\documentclass[a4paper,11pt]{article}
\usepackage[utf8]{inputenc}
\usepackage[table,xcdraw]{xcolor}
\usepackage{hyperref}
\usepackage{mathrsfs,amsmath, amssymb,relsize}
\usepackage{multirow,multicol,array}
\usepackage{array}
\usepackage{graphicx}
\usepackage{subcaption}
\usepackage{lipsum}
\usepackage{booktabs}
\usepackage{lscape}
\usepackage{comment}
\usepackage{soul}
\usepackage{tikz}
\usepackage[a4paper, total={6in, 8in}]{geometry}
\usepackage{tabularx}
\usepackage{cleveref}
\usepackage{authblk}

\usepackage{subcaption}
\usepackage{ulem}

\hypersetup{
    colorlinks=true,
    linkcolor=green,
    filecolor=magenta,      
    urlcolor=green,
    citecolor=green,
}

\newcommand\remove[1]{}

\usepackage{color}
\definecolor{darked}{rgb}{0.5,0,0}
\definecolor{green}{rgb}{0,0.8,0}
\definecolor{darkgreen}{rgb}{0.3,0.7,0.3}
\definecolor{darkblue}{rgb}{0.0314,0.2745,0.5477}
\definecolor{darkred}{rgb}{0.6350, 0.0780, 0.1840}
\definecolor{red}{rgb}{1, 0.180, 0.1840}
\definecolor{clearblue}{rgb}{0.3124,0.6092,0.7974}
\definecolor{fadedblue}{rgb}{0.5529,0.7556,0.8680}
\definecolor{blue}{rgb}{0,0.3,1}
\definecolor{orange}{rgb}{ 0.9,0.6,0.1253}
\definecolor{darkorange}{rgb}{0.8,0.4,0.05}
\definecolor{lightblue}{rgb}{0.1,0.1,1}
\definecolor{ocre}{rgb}{0.6, 0.6, 0.3}

\title{A Python-based flow solver for  numerical simulations \\ using an immersed boundary method on single GPUs}

\author[1]{M. Guerrero-Hurtado}
\author[1]{J. M. Catalán\footnote{Email address for correspondence: \href{mailto:jcatalan@ing.uc3m.es}{jcatalan@ing.uc3m.es}}}
\author[2]{M. Moriche}
\author[3]{A. Gonzalo}
\author[1]{O. Flores}
\affil[1]{Department of Aerospace Engineering, Universidad Carlos III de Madrid, Spain}
\affil[2]{Institute of Fluid Mechanics and Heat Transfer, Technische Universität Wien, Austria}
\affil[3]{Department of Mechanical Engineering, University of Washington, Seattle, WA, United States}

\date{}

\newcommand{\orangesquare}{\raisebox{2pt}{\tikz{\draw[-,darkorange,solid,line width = 5pt](0,0) -- (4mm,0);}}}
\newcommand{\bluesquare}{\raisebox{2pt}{\tikz{\draw[-,clearblue,solid,line width = 5pt](0,0) -- (4mm,0);}}}

\newcommand{\dasheddarkergreenline}{\raisebox{2pt}{\tikz{\draw[-,darkgreen,dashed,line width = 2pt](0,0) -- (3mm,0);}}}
\newcommand{\dashedclearblueline}{\raisebox{2pt}{\tikz{\draw[-,clearblue,dashed,line width = 2pt](0,0) -- (3mm,0);}}}

\begin{document}

\maketitle

\begin{abstract}
We present an efficient implementation for running three-dimensional numerical simulations of fluid-structure interaction problems on single GPUs, based on Nvidia CUDA through Numba and Python.
The incompressible flow around moving  bodies is solved in this framework through an implementation of the Immersed Boundary Method tailored for the GPU, where different GPU grid architectures are 
exploited to optimize the overall performance.
By targeting a single-GPU, we avoid GPU-CPU and GPU-GPU communication bottlenecks, since all the simulation data is always in the global memory of the GPU.
We provide details about the numerical methodology, the implementation of the algorithm in the GPU and the memory management, critical in single-GPU implementations.
Additionally, we verify the results comparing with our analogous CPU-based parallel solver and assess satisfactorily the efficiency of the code in terms of the relative computing time of the different operations and the scaling of the CPU code compared to a single GPU case.
Overall, our tests show that the single-GPU code is between 34 to 54 times faster than the CPU solver in peak performance (96-128 CPU cores).
This speedup mainly comes from the change in the method of solution of the linear systems of equations, while the speedup in sections of the algorithm that are equivalent in the CPU and GPU implementations is more modest (i.e., $\times 1.6-3$ speedup in the computation of the non-linear terms). 
Finally, we showcase the performance of this new GPU implementation in two applications of interest, one for external flows (i.e., bioinspired aerodynamics) and one for internal flows (i.e., cardiovascular flows),  demonstrating the strong scaling of the code in two different GPU cards (hardware).%\red{for both single and double precision.}

\end{abstract}
\section{Introduction} \label{sec:introduction}

Over the past few decades the fluid dynamics community has dealt with the significant challenge of conducting numerical simulations of flows around bodies in motion and undergoing deformation, particularly those with intricate geometries. 
Examples range from high-speed flows \cite{young2013numerical,ferrer2015cfd} to a large variety of low-to-moderate Reynolds number flows, including the sedimentation of particles in a fluid \cite{balachandar2010turbulent,brandt2022,uhlmann2014sedimentation,moriche2023clustering}, the complex blood motion present in cardiovascular flows \cite{meschini2018flow,garcia2021demonstration} or the unsteady flow dynamics at play in flying insects \cite{berman2007energy}, micro-air vehicles \cite{gonzalo2018} or aquatic living beings \cite{becker2015hydrodynamic}, among others.
For an extended period, significant effort has been dedicated to modeling and simulating such systems.

Two main approaches have been traditionally employed to address the treatment of moving bodies submerged in a fluid. The first consists of solving the fluid equations with a grid that adapts to the curvature of the body boundary (body-conformal grid methods), where the no-slip boundary condition can be easily imposed at the surface of the body.
The second employs a grid for the fluid that is independent on the shape and position of the submerged body (non-body conformal grid methods). 
In this case, the no-slip boundary condition at the surface of the body is imposed differently, through an additional source term in the flow equations acting only at grid points in the vicinity of the body.
Techniques using this strategy conform the so-called immersed boundary methods (IBM), also known as embedded boundaries or immersed interfaces.

Several different body conformal techniques exist and have proved successful in many applications \cite{tezduyar2001finite,haeri2012application}.
The main idea is that either the whole mesh or a part of it moves locally with the body.
A typical example is a technique that employs two different non-overlapping grids: a static mesh and a sliding mesh that moves with the body at each time step.
This strategy is usually used in rotor configurations, where the sliding mesh moves according to the angular motion of the rotor blades \cite{steijl2008sliding, mcnaughton2014simple}.
In this case, since one mesh is moving with respect to the other, an additional step is required to identify the connection nodes between these two grids at every time step, and perform interpolations if needed.
Another widely-used technique consists of having local mesh deformations as the body moves, typically used in fluid-structure interaction (FSI) problems where the body flexibility is not negligible \cite{sanchez2016towards,economon2016su2,yang2018aerodynamic}.
This method only uses a single mesh and thus avoids the need for identifying connection nodes and interpolating, as in the previous case.
However, the mesh quality might be compromised if the body undergoes motions with very large amplitudes.
Finally, there are also alternative methods in which re-meshing is needed at each step \cite{ramamurti2001simulation}.
For instance, the Arbitrary Lagrangian-Eulerian (ALE) formulation is an example of this strategy
\cite{baum1998numerical,hirt1974arbitrary,donea1982arbitrary,takashi1992arbitrary,bazilevs2008isogeometric,souli2013arbitrary}.

Alternatively, immersed boundary methods have been thoroughly investigated over the last decades \cite{mittal2005immersed,griffith2020immersed}  and have gained considerable popularity with the growth of computational resources \cite{verzicco2023immersed}.
Originally introduced by Peskin \cite{peskin1972flow,peskin1977numerical,peskin2002immersed}, the IBM methods define two independent grids: a fixed grid for the fluid (the Eulerian grid), and a moving grid that discretizes the body (the Lagrangian grid). 
The key point in any IBM is the coupling between the variables defined in the Eulerian and the Lagrangian grids to account for the interaction between the fluid and the body.
This is performed by means of regularized delta functions \cite{peskin1972flow,Roma1999adaptive,uhlmann2005}, whose definition is crucial in terms of smoothness and conservation properties of the solution. 
Additionally, there are other IBM techniques that use the so-called feedback forcing methods \cite{goldstein1993modeling} and penalty methods \cite{specklin2018sharp}, where the local forcing term is computed differently.

One of the clear advantages behind this family of methods lies in the elimination of the complex and computationally expensive task of regenerating the fluid mesh at each time step as the immersed body undergoes motion or deformation, as it happens in body-fitted grid methods.
Another advantage is its suitability for massive parallelization, facilitated by the straightforwardness of the Cartesian grids usually employed for the fluid. 
This simplicity enables a seamless decomposition of the computational domain in subregions. 
Nonetheless, caution is required during the interpolation and spreading processes between the fluid mesh and the body mesh, and vice versa. 
This is particularly crucial as individual segments of the body surface may exist in diverse subregions of the parallelized domain \cite{uhlmann2004simulation,wang2013parallel,spandan2017parallel}. 
Additionally, solving linear systems may pose extra challenges in the context of parallelization \cite{myllykoski2018solving,jodra2017solving}.

Immersed boundary method implementations have historically been developed for CPU parallel architectures, typically used in high-performance computing clusters, given their associated high computational cost \cite{constant2017immersed,yildirim2013parallel,wang2013parallel,krause2017incompressible}.
For instance, our group has developed TUCAN \cite{moriche2017numerical,gonzalo2018aerodynamic}, an extensively validated massive parallel solver that has proven to be effective in simulating moving bodies immersed in a fluid \cite{moriche2017,arranz2022flow,martinez-muriel2023}. 
In TUCAN, the Navier-Stokes equations are solved using a fractional step method, where the linear systems are tackled iteratively using HYPRE \cite{falgout2002hypre}. 
Additionally, parallelization is achieved via a three-dimensional domain decomposition utilizing MPI \cite{walker1996mpi}.
The numerical approach proposed in this work is very similar to TUCAN, although some relevant differences in terms of parallelization, target architectures, solution of the linear systems and overall performance are present and will be detailed in the following sections.

In this framework, the rapid progress of technology in recent years has paved the way for Graphical Processing Units (GPUs) to become a prominent player in the world of high-performance computing. 
The advancements in GPU technology, driven by the rise of artificial intelligence (AI), have significantly impacted the field of computational fluid dynamics (CFD) \cite{niemeyer2014recent, afzal2017parallelization, cary2021cfd}, thanks to their highly parallel architecture, fast processing capabilities and exponentially increasing memory capacity.
In the last years, the fluid dynamics community has leveraged GPU devices to manage the high computational load and complexity associated to these simulations.
On the one hand, some researchers have increasingly focused on developing strategies for accelerating numerical simulations using GPUs. 
Some of the effort dedicated to these strategies have centered on improving the code's performance using programming models such as OpenACC \cite{openacc}. This approach usually provides a quite reasonable acceleration, without needing an intensive programming effort \cite{raj2023gpu}.
Other authors, alternatively, have invested time and resources in performing an acceleration tailored to the GPU, based on their traditional multi-CPU parallel solvers.
For legacy reasons, many of these developments have focused on targeting multi-GPU architectures, typically using the Compute Unified Device Architecture from Nvidia \cite{cuda} (CUDA).
Such is the case of the porting described in \cite{viola2022fsei} for the fluid-structure-electrophysiology interaction (FSEI) in the left heart and \cite{zhu2018afid,vela2021low} for wall-bounded turbulent flows.

Related to numerical solvers using the immersed boundary method in GPUs, several alternatives have been proposed in the research community. One of them is the work by Di \& Ge \cite{di2015simulation}, where they present a modified algorithm of the IBM proposed by Uhlmann \cite{uhlmann2005}, accelerated on the GPU. In this work, they claim improvements mainly in terms of the accuracy of the boundary-fluid interface. To facilitate the GPU implementation, they treat all the terms in the Navier-Stokes equations explicitly.
In other works such as  Viola et al. \cite{viola2022fsei}, the authors present a multi-GPU acceleration of the whole FSEI problem, aiming to address the complete phenomenology of cardiovascular flows. They implement the no-slip boundary condition on the wet heart tissues through an IBM based on the moving least square approach \cite{vanella2009moving,de2016moving} and they perform the GPU parallelization using CUDA kernels, which are functions executed by the GPU threads individually, providing a significant speed-up compared to the CPU version of the code.

Alternatively, special attention has been paid to the acceleration of codes based on the Lattice-Boltzmann Method (LBM), due to the versatility it provides and the relative simplicity of the parallelization given its explicit character. Examples are the work by Rinaldi et al. \cite{rinaldi2012lattice}, where the authors focus on optimizing the data transfer within the GPU device, making use of the shared memory to maximize memory bandwidth; or the study by Valero-Lara et al. \cite{valero2015accelerating}, where they present an implementation of LBM coupled with IBM on heterogenous architectures, centering on the details of the parallelization via user-defined CUDA kernels.
Other authors such as Ames et al. \cite{AMES2020101153}, on the other hand, propose a multi-GPU IBM-LBM coupling for cell-resolved hemodynamic simulations featuring fluid-structure interaction through a finite-element solver for the structure. With that purpose, they focus on describing the complex parallelization framework and report performance, scaling and data transfer comparison between the GPU and CPU implementations.

The increased power and efficiency of modern GPUs now allow some of the aforementioned tasks to be performed more effectively without needing a multi-GPU framework. 
The parallel processing capabilities, improved memory bandwidth, and enhanced architecture of contemporary GPUs enable them to handle extensive calculations and data processing directly on the GPU. 
This eliminates the need for frequent data transfers between the GPU and CPU or other GPUs, thereby avoiding the communication bottlenecks that traditionally hinder performance. 
Consequently, the increase in relative performance due to this streamlined, on-GPU processing further enhances the efficiency and feasibility of conducting numerical simulations on a single-GPU,
allowing to perform even parametric sweeps on moderately large problems.
In this regard, some works exploiting single-GPU architectures can be found. %for two-dimensional applications. 
One of them is cuIBM \cite{layton2011cuibm}, a two-dimensional code that implements the IBM proposed by Taira \& Colonius \cite{taira2007immersed} and employs GPU-accelerated basic linear algebra subroutines from CUDA for handling sparse matrices.
Another clear example is the work by Wu et al. \cite{wu2019gpu}, which concentrates on targeting a single GPU implementation of the coupling between LBM and IBM, using the work of Tölke \cite{tolke2010implementation} as reference. In particular, they design two different kernels: the spreading kernel, featuring a fluid mesh centric arrangement (or Eulerian mesh centric arrangement) with a three-dimensional grid architecture, and the interpolation and boundary force computation kernel, that showcases a one-dimensional body (Lagrangian) mesh centric arrangement for the calculations.

In this work, we conveniently exploit the architecture of graphical processing units, presenting a single-GPU parallel implementation for running three-dimensional simulations of the flow around moving bodies with a significant performance boost, using a direct forcing immersed boundary method.
By targeting a single-GPU, we leverage the high memory bandwidth of recent GPU cards and hence avoid communication bottlenecks between devices that a multi-GPU implementation needs to deal with. 
In the current implementation, full parallelization is achieved by employing different strategies tailored to the GPU, such as the design of user-defined kernels, enabling efficient computations and memory management, as well as the utilization of highly optimized libraries from CUDA, significantly accelerating the solution of the linear systems. 
A clear example is the use of the CuFFT library \cite{cuda}, suitable for configurations exhibiting periodic directions, something common in a wide range of scientific problems \cite{uhlmann2014sedimentation,moriche2023clustering,vela2021low}.
We employ a high-level, general-purpose programming language like Python \cite{van1995python}, offering an efficient and user-friendly development environment, together with CUDA.
The access to the CUDA features is provided through the Numba library \cite{numbacuda}, an open-source just-in-time compiler that translates a subset of Python code into fast machine code using LLVM (Low-Level-Virtual-Machine), as well as through CuPy \cite{cupy}, an open source library for GPU-accelerated computing that shares the same API set as NumPy \cite{harris2020array} and SciPy \cite{2020SciPy-NMeth}.

The rest of the manuscript is structured as follows: section \ref{sec:methodology} describes the methodology, containing the governing equations, numerical approach and GPU implementation; section \ref{sec:verification} details the verification of the results; section \ref{sec:code-performance} presents the performance of the code, section \ref{sec:applications} showcases some applications and finally section \ref{sec:conclusions} draws some conclusions.

%%%%%%%%%%%%
%%%%%%%%%%%%
%%%%%%%%%%%%%%%%%%%%%%%%%%%%%%%%%%%%%%%%%%%%%%%%%%%%%%%%%%%%%
%%%%%%%%%%%%%%%%% Methodology
%%%%%%%%%%%%%%%%%%%%%%%%%%%%%%%%%%%%%%%%%%%%%%%%%%%%%%%%%%%%%
\section{Methodology}\label{sec:methodology}
%%%%%%%%%%%%%%%%%%%%%%%%%%%%%%%%%%%%%%%%
%%%%%%%%%%% Governing equations
%%%%%%%%%%%%%%%%%%%%%%%%%%%%%%%%%%%%%%%%
\subsection{Governing equations and numerical scheme}
\label{sec:GovEqNumScheme}

We consider a body moving in a Newtonian fluid with constant kinematic viscosity $\nu$ and density $\rho$. The fluid velocity and pressure are governed by the Navier-Stokes equations for an incompressible flow, which written in dimensionless form are 
\begin{subequations} \label{eq:NS}
\begin{align}
      \nabla \cdot \mathbf{u} \, = & \,0,  \label{eq:NSa}\\
      \frac{\partial \mathbf u}{\partial t}  + (\mathbf u\cdot \nabla) \mathbf u \, = & - \nabla p + \mathrm{Re}^{-1} \nabla^2 \mathbf u + \mathbf f, \label{eq:NSb} \\
      \mathbf{u} = & \,\mathbf{U}^d \quad \text{at the surface of the body}. \label{eq:NSc}
\end{align}
\end{subequations}
These equations are made dimensionless using the fluid density $\rho$, a characteristic velocity $U_c$ and the characteristic length $L_c$. 
Hence, $\mathbf u$ and $p$ are the dimensionless velocity and pressure, $\mathrm{Re} = U_c L_c/\nu$ is the Reynolds number and $\mathbf{U}^d$ is the velocity of the surface of the body.
As described below, the volumetric force $\mathbf f$ is the Immersed Boundary Method (IBM) forcing term, which enforces the no-slip boundary condition at the surface of the body (i.e., eq. \ref{eq:NSc}). 
Additionally, the problem requires initial conditions and boundary conditions for both $\mathbf{u}$ and $p$.

In order to solve eqs. \eqref{eq:NS}, we use a fractional step method \cite{brown2001accurate}, where \eqref{eq:NSa} is enforced by decomposing $\mathbf u$ in a non divergence-free term $\mathbf u^*$ (hereafter intermediate velocity) and an irrotational term equal to the gradient of a potential $\phi$ (hereafter pseudo-pressure).
The time integration is performed with a self-starting low-storage semi-implicit Runge-Kutta method with three stages \cite{rai1991direct}.  

The spatial discretization is
performed with a Cartesian uniform, staggered grid.  
In the staggered grid, scalar variables (i.e, $p$) are defined at the center of the cells, while the velocity components are defined at the center of cell faces.
The spatial operators are discretized with second-order, central finite-differences.
To keep the global order of accuracy of the method, we employ ghost points to enforce the boundary conditions at the boundaries of the computational domain.  
This allows using the same central scheme for the spatial derivatives for internal and boundary grid points. 

The presence and motion of the solid body is modelled with the IBM proposed by Uhlmann \cite{uhlmann2005}. Hence, the volumetric force $\mathbf{f}$ is added to the right-hand-side of the momentum equation \eqref{eq:NSb}, to force the fluid velocity to match the velocity of the body's surface ($\mathbf{U}^d$) in the vicinity of the body.
%
%We use the direct forcing formulation of the IBM proposed by Uhlmann \cite{uhlmann2005}.
%
The body's surface is discretized with $N_\eta$ evenly spaced grid points (i.e, the  Lagrangian grid), with approximately the same grid spacing used for the staggered grid of the fluid (i.e., the Eulerian grid).
The interpolation of velocities and forces between the Eulerian and Lagrangian grid is tackled using the discrete approximation of Dirac's regularized delta function, $\delta_h$ \cite{peskin2002immersed,Roma1999adaptive}.

The equations corresponding to the $n$-th Runge-Kutta stage, including the volumetric force that couples fluid and solid interactions ($\mathbf f$ and $\mathbf F$, in the Eulerian and Lagrangian frames, respectively), read:

\begin{subequations}\label{rksubstep:main}
\begin{align}
\begin{split}
    & \mathbf u^e  =  \mathbf u^{n-1} +  \Delta t [ (\alpha_n + \beta_n)\mathrm{Re}^{-1} \nabla^2 \mathbf u^{n-1} - (\alpha_n + \beta_n) \nabla p^{n-1} \\& \hspace*{2cm}  -      \gamma_n[(\mathbf u \cdot \nabla)\mathbf u]^{n-1} - \zeta_n[(\mathbf u \cdot \nabla)\mathbf u]^{n-2} ],   \label{rksubstep:a}
\end{split}\\%[-0.5cm]
\begin{split}
  &  \mathbf U^e(\mathbf X_{\eta}) = \sum_{\xi\in N_{\xi}} \mathbf                u^e (\mathbf{x}_{\xi})\delta_h(\mathbf x_{\xi} - \mathbf X_{\eta})h^3,  \label{rksubstep:b}
\end{split}\\%[-0.5cm]
\begin{split}
  &  \mathbf{F} (\mathbf X_{\eta}) = \frac{\mathbf{U}^d (\mathbf X_{\eta}) -  \mathbf U^e (\mathbf X_{\eta}) }{\Delta t},  \label{rksubstep:c}
\end{split}\\%[-0.5cm]
\begin{split}
 &   \mathbf f (\mathbf x_{\xi}) = \sum_{\eta\in N_\eta} \mathbf F (\mathbf X_{\eta}) \delta_h(\mathbf x_\xi - \mathbf X_{\eta})\Delta V_{\eta},    \label{rksubstep:d}
\end{split}\\%[-0.5cm]
\begin{split}
& \nabla^2 \mathbf u^* - \frac{\text{Re}}{\beta_n \Delta t} \mathbf u^* = -              \frac{\text{Re}}{\beta_n \Delta t} (\mathbf u^e - \Delta t \beta_n \mathrm{Re}^{-1}      \nabla^2 \mathbf u^{n-1} + \Delta t \mathbf f),
\label{rksubstep:e}
\end{split}\\%[-0.5cm]
\begin{split}
& \nabla^2 \phi = \frac{\nabla \cdot \mathbf u^*}{(\alpha_n + \beta_n) \Delta t},
\label{rksubstep:f}
\end{split}\\%[-0.5cm]
\begin{split}
& \mathbf u^n = \mathbf u^* - (\alpha_n + \beta_n) \Delta t \nabla \phi,
\label{rksubstep:g}
\end{split}\\%[-0.5cm]
\begin{split}
& p^n = p^{n-1} + \phi - \frac{\beta_n \Delta t}{\text{Re}} \nabla^2 \phi.
\label{rksubstep:h}
\end{split}
\end{align}
\end{subequations}
We use capital letters (i.e., $\mathbf{U}^d, \mathbf{F}, \mathbf{X}$) to denote variables defined in the Lagrangian grid, and lower case letters (i.e., $\mathbf{u}, p, \mathbf{f}, \mathbf{x}$) to refer to variables defined in the Eulerian grid.  
In particular, $\mathbf u^{e}$ and $\mathbf U^e$ correspond to an explicit velocity estimation, used for the calculation of the IBM forcing term ($\mathbf f$ and $\mathbf F$),  $\eta$ is the index running over the total number of Lagrangian points $N_\eta$, $\xi$ is the index running over the $N_\xi$ Eulerian points neighbouring $\mathbf{X}_\eta$, $h$ is the grid spacing in the Eulerian grid, $\Delta t$ is the time step and $\Delta V_{\eta}$ is the volume associated to each Lagrangian point. 
Finally, $\alpha_n, \beta_n, \gamma_n$, and $\zeta_n$ are the coefficients of the Runge-Kutta scheme.
The interested reader is refered to \cite{uhlmann2005} for more details.

From the point of view of the implementation of this algorithm into a solver, the operations described in \eqref{rksubstep:main} can be organized in three groups: Eulerian operations, Lagrangian operations, and solution of linear systems of equations.  
In the first group, Eulerian operations, we have the evaluation of the right-hand-sides of equations \eqref{rksubstep:a}, \eqref{rksubstep:e}, and  \eqref{rksubstep:g}, together with the evaluation of the correction steps (eqs. \ref{rksubstep:g} and \ref{rksubstep:h}). These are explicit operations performed in the Eulerian grid, involving neighbouring grid-points, sums and products. 
In the second group, Lagrangian operations, we have the evaluation of the IBM forcing $\mathbf{f}$ from the estimated velocity field, $\mathbf{u}^e$  (eqs.  \ref{rksubstep:b}, \ref{rksubstep:c} and \ref{rksubstep:d}). This involves the evaluation of $\delta_h(\mathbf{x}_\xi - \mathbf{X}_{\eta})$ for each Lagrangian point, both in eq. \eqref{rksubstep:b} and \ref{rksubstep:d}. 
The third group includes the Helmholtz \eqref{rksubstep:e} and Poisson \eqref{rksubstep:f} problems. These two equations require solving four linear systems of equations with $N$ unknowns each, where $N$ is the number of cells in the Eulerian grid.
The solution of the linear systems (particularly the Poisson problem) is usually the most computationally demanding part of the solver. 
Consequently, the choice of the method used to solve them (i.e., direct or iterative methods, choice of preconditioners, etc.) is of paramount importance. 

%%%%%%%%%%%%%%%%%%%%%%%%%%%%%%%%%%%%%%%%
%%%%%%%%%%% GPU Implementation
%%%%%%%%%%%%%%%%%%%%%%%%%%%%%%%%%%%%%%%%
\subsection{Design considerations} \label{sec:DesignConsiderations} 

\begin{figure} [t!]
    \centering
    %\hspace*{-0.5cm}
    \includegraphics[width= \textwidth]{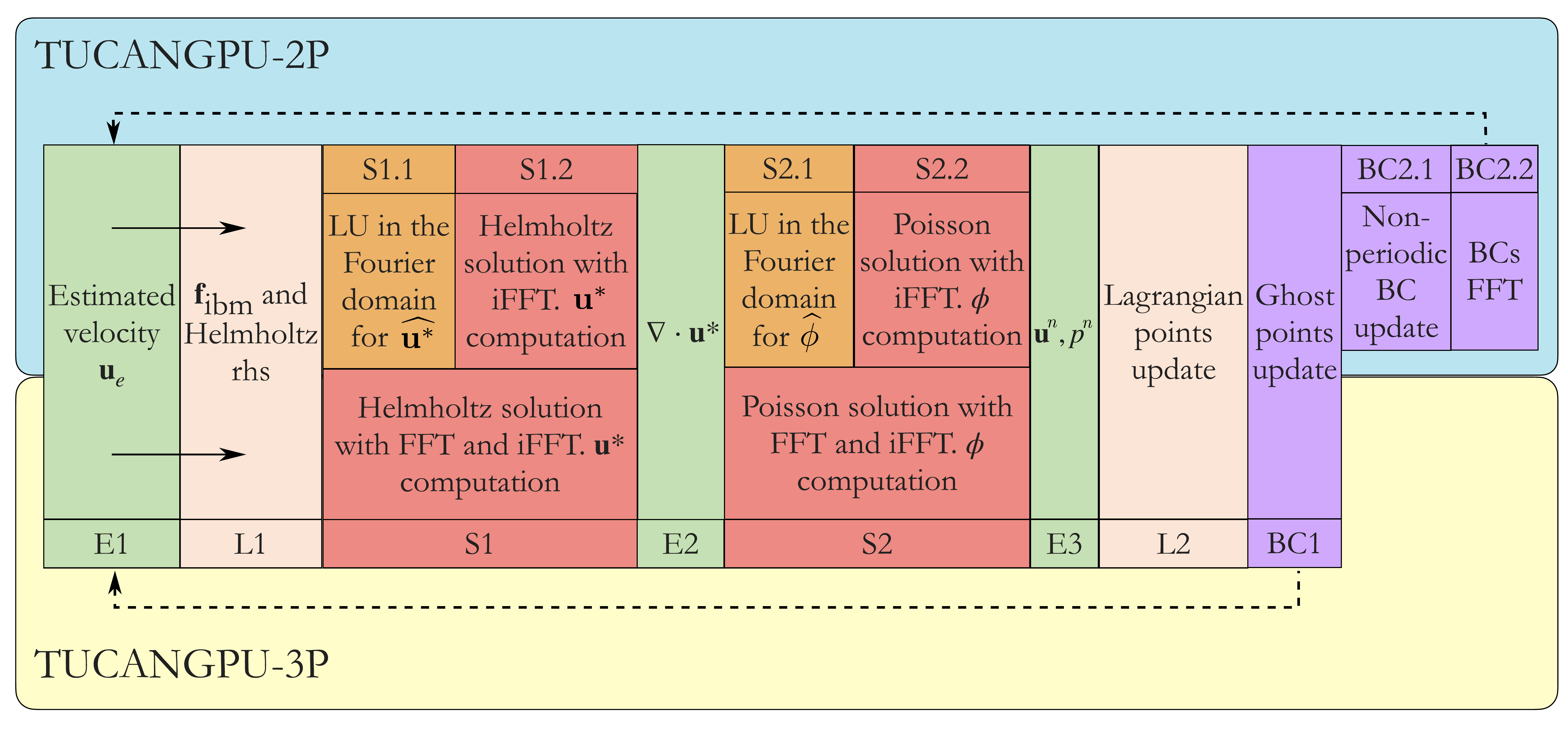}
    \caption{Unified flowchart of the operations performed by the different TUCANGPU branches, in a single stage of the Runge-Kutta 3 (RK3) temporal scheme. The upper stream (blue background) refers to TUCANGPU-2P and the lower (yellow background) to TUCANGPU-3P. Eulerian grid kernels are coloured by green and Lagrangian kernels by light orange. Kernels coloured with intense orange are those related with a two-dimensional GPU grid architecture, only present in the TUCANGPU-2P branch. Finally, red boxes account for direct and inverse Fourier Transforms. Operations are executed from left to right.
    When the last operations of the RK3 stage are completed (BCs, in purple), the process repeats again, as indicated by the dashed line.
    Note that each box is associated to a specific label, for instance, the computation of the estimated velocity $\mathbf{u}_e$ (first green box) corresponds to the E1 label.
    }
    \label{fig:flowchart}
\end{figure}

We choose to develop our single-GPU solver (TUCANGPU) in Python, accessing the NVIDIA CUDA Toolkit \cite{cuda} through Numba \cite{numbacuda} for programming the kernels, and using CuPy \cite{cupy} to access GPU-accelerated libraries. 
Understanding the GPU architecture and the CUDA framework is necessary to ensure an efficient implementation of eqs. \eqref{rksubstep:main} in TUCANGPU.

Within the CUDA framework, the operations executed on the GPU are coded via kernels, which contain the operations launched by each individual thread.
Threads are the fundamental units of execution on the GPU. 
CUDA threads are organized hierarchically into blocks, which are further grouped into a grid (i.e., the GPU grid).
Each thread block operates independently and asynchronously on an available streaming multiprocessor (SM) within the GPU. 
All thread blocks have access to the {\it global} memory of the GPU card, which allows communication between the different thread blocks. This global memory is not in the SM, and accessing it from individual threads is not efficient. 
Each thread block has its own {\it shared} memory (i.e., on the SM, fast), accessible from all its threads. 
Each of these threads has its own {\it local} memory (faster, limited), which cannot be accessed from other threads.  
Finally, the communication between the CPU memory and the GPU memory is usually not efficient, presenting a common bottleneck in heterogeneous CPU-GPU codes \cite{gregg2011data}.

Taking all these into consideration, the following principles have been followed in the design of the implementation of eqs. \eqref{rksubstep:main} into TUCANGPU: 
\begin{enumerate} 
\item Minimize CPU-GPU communication. Since we are targeting a single-GPU implementation, there is no need to access the CPU memory to communicate data between different GPU threads. CPU is only tasked with the I/O operations. Everything else is done inside the GPU card. 
\item Optimize memory access in the GPU card. Three different types of kernels will be designed, to optimize the data transfer between global, shared and local memories for the Eulerian/Lagrangian operations and deal with the solution of the linear systems of equations. Care will be taken to deal with the asynchronous operation of the thread blocks, avoiding race conditions. 
\item Critical review of the numerical methods used to solve eqs. \eqref{rksubstep:main}, leveraging GPU-accelerated libraries to maximize code performance and minimize coding effort. 
This has lead to a different strategy to solve the linear systems in the GPU implementation, using the FFT libraries provided by CuPy (see details in section \ref{sec:LinSystems}). 

\end{enumerate} 

%%%%%%%%%%%%%%%%%%%%%%%%%%%%%%%%%%%%%%%%%%%%%%%%%%%%%%%%%%%%%%%
\subsection{TUCANGPU implementation, workflow and kernels}

In TUCANGPU the CPU is primarily responsible for command execution and handling Input-Output operations using the HDF5 library \cite{hdf5}. 
All other operations are performed inside the GPU, which requires defining all buffers, auxiliary variables and parameters in the {\it global} memory of the GPU at the initialization phase of the solver. 
To minimize read/write operations from/to {\it global} memory, GPU kernels are designed to integrate as many operations as possible before requiring communication with the {\it global} memory. 
This design approach implies the reduction of the number of kernel calls during the execution of each Runge-Kutta stage. 

Figure \ref{fig:flowchart} presents the workflow used in TUCANGPU to perform a substep of the Runge-Kutta time integrator. 
The different subroutines involved in the workflow are color-coded into four groups, have a specific label: Eulerian operations (E1, E2, E3), Lagrangian operations (L1, L2), solution of systems of linear equations (S1, S2, S1.1, S1.2, S2.1, S2.2), and set up of boundary conditions (BC1, BC2.1, BC2.2).  
Note that the first three groups correspond to the three types of operations described at the end of section \ref{sec:GovEqNumScheme}.
Eulerian and Lagrangian operations are coded into ad-hoc kernels using Numba, while the solution of the linear systems of equations and the subroutines to set up boundary conditions are implemented using CuPy functions and libraries.

It is important to emphasize that the workflow of TUCANGPU in fig. \ref{fig:flowchart} features two versions for the subroutines associated to the solution of the linear systems of equations and the set up of boundary conditions. As discussed in section \ref{sec:LinSystems} below, they correspond to two different versions of the solver: TUCANGPU-3P which solves eqs. \eqref{rksubstep:main} for a triply-periodic computational domain, and TUCANGPU-2P which solves eqs. \eqref{rksubstep:main} for a double periodic computational domain, with non-periodic (for instance, inflow/outflow) boundary conditions in the third direction. 
Although seemingly restrictive, having these two configurations allows to cover a great number of scientific problems of interest, such as particle-laden flows, cardiovascular flows, external aerodynamics, wall-bounded flows, etc.

\begin{figure}[t!]
    \centering
    \includegraphics[width= 0.9\textwidth]{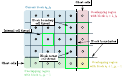}
    \caption{Two-dimensional grid featuring a block-type domain decomposition. 
    Different kinds of thread are highlighted with green borders depending on the information they need to load beyond the physical boundary of the block.
    For simplicity, only the pressure grid points are represented, with points inside the physical domain of the block filled in black and those outside (ghost cells) filled in white.
    In this work we extend this idea to the grid of the velocity components, not present in this sketch, as well as to the third dimension.
    }
    \label{fig:block_diagram}
\end{figure}

\subsubsection{Kernels for Eulerian operations}
\label{sec:EuKernels}

These kernels handle the evaluation of the right-hand-sides of eqs. (\ref{rksubstep:a}, 
%\ref{rksubstep:e}, 
\ref{rksubstep:f}, \ref{rksubstep:g}, \ref{rksubstep:h}, \ref{rksubstep:e}). 
To efficiently parallelize these tasks, each thread is tasked with the computation of the right-hand-side for a single Eulerian cell. 
The presence of divergence and gradient operations, which are discretized using second-order finite differences approximations, implies that each thread requires information from neighbouring cells in all three directions. 
Hence, thread blocks are defined by dividing the 3D Cartesian space into rectangular subdomains (i.e., block domain decomposition), and the {\it shared} memory of each thread block requires storing fluid variables for the interior cells, and fluid variables for the cells neighbouring the block's subdomain (i.e., ghost cells). This is sketched in figure \ref{fig:block_diagram} for a 2D block decomposition, showing the ghost cells associated to a given block ($i_b,j_b$). 

To ensure efficient memory accesses, the block's {\it shared} memory is allocated at the beginning of each kernel execution.
Specifically, each thread is tasked with copying from {\it global} to {\it shared} memory the required fluid variables on its assigned cell, and the fluid variables of any adjacent ghost cell. 
This process is described in figure \ref{fig:block_diagram}, where interior/boundary/corner cells need to load into {\it shared} memory the data in the green boxes. 
Then, the kernel's calculations are performed using only data stored in the {\it shared} memory. 
Once completed, each thread updates the {\it global} memory with the computed right-hand-side before concluding its execution.

To minimize the communication between {\it global} and {\it shared} memory, we define three Eulerian kernels: 
\begin{itemize}
    \item \textbf{E1}, {\it Explicit Terms}: this kernel computes $\mathbf u^e$ (see eq. \ref{rksubstep:a}), and the right-hand-side of eq. (\ref{rksubstep:e}) without the IBM forcing term $\mathbf f$, which is added later (see \ref{sec:LagKernels} below). 
    Each thread computes the sum of terms for a single cell within the Eulerian grid, collecting the required fluid variables from the {\it global} memory using the procedure described in the previous paragraph. 
    \item \textbf{E2}, {\it Divergence Calculation}: the input data for this kernel are the three velocity components of $\mathbf{u}^*$ (plus ghost cells).
    After loading the required data into {\it shared} memory, each thread computes $\nabla \cdot \mathbf{u}^*$ and saves it in the {\it global} memory. 
    \item \textbf{E3}, {\it Pseudo-pressure projection}: this kernel computes $\mathbf u^{n}$ and $p^n$. 
    Each thread is tasked with correcting the velocity with the spatial gradient of the pseudo-pressure (see eq. \ref{rksubstep:g}), and with correcting the pressure with the (already-known) Laplacian of $\phi$ (see eq. \ref{rksubstep:h}).
    Hence, each thread loads $\phi$ and $\nabla \cdot \mathbf u^*$ from the {\it global} memory into the {\it shared} 
    memory (plus ghost cells), then computes the required terms, and finally copies $\mathbf u^{n}$ and $p^n$ into the {\it global} memory.  
\end{itemize}

\subsubsection{Kernels for Lagrangian operations}
\label{sec:LagKernels}

\begin{figure} [t!]
    \centering
    \includegraphics[width= 0.9\textwidth]{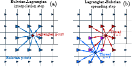}
    \caption{Interpolation (a) and spreading (b) scheme for the computation of the immersed boundary volumetric force $\mathbf f$. For the sake of clarity, a 3 points interpolation scope per direction has been considered for this scheme. In (b), Eulerian points coloured with red are those collecting information only from the red Lagrangian marker (with index $\eta=1$), while the point $\eta=2$ (blue) spreads its $\mathbf F$ over those Eulerian points coloured with blue and purple. Eulerian points coloured with purple are those collecting information from both Lagrangian markers.}
    \label{fig:force_lag}
\end{figure}

The Lagrangian operations include the interpolation/spreading operations associated with the IBM, and the update of the position and velocity of the body's surface at the end of each time-step. 
All these operations have in common that there is no communication between adjacent Lagrangian points, enabling the parallelization of these tasks using a one-dimensional GPU grid architecture that comprises the $N_\eta$ Lagrangian points discretizing the body's surface.
Moreover, since the number of Eulerian points involved in the operations of each Lagrangian points is small (i.e., the scope of the delta functions is usually small, with 4-5 grid points in each dimension, see \cite{peskin1972flow,Roma1999adaptive}), these operations can be performed using only the {\it local} memory of the thread. 

Below, we detail the two Lagrangian kernels incorporated in the code:
\begin{itemize}

    \item \textbf{L1}, {\it Forcing Term}: this kernel computes the immersed boundary force in the Eulerian mesh, $\mathbf{f}$, and adds it to the buffer in the {\it global} memory that stores the terms of the right-hand-side of eq. \eqref{rksubstep:e} already computed by the Eulerian kernel E1. 
    The input data for this Lagrangian kernel are the estimated velocity $\mathbf{u}^e$, the position and velocity of the Lagrangian points, $\mathbf{X}_{\eta}$ and $\mathbf{U}^d$. 
    At the beginning of the kernel, each thread is tasked with loading the 3 velocity components of its influence region into its {\it local} memory, as sketched in figure \ref{fig:force_lag}a. 
    Then, each thread performs all the necessary evaluations of the {\it delta} functions, storing them in the {\it local} memory. 
    These evaluations of the delta functions will be used both in the interpolation (eq. \ref{rksubstep:b}) and spreading (eq. \ref{rksubstep:d}) operations. 
    With this data, the thread proceeds with the interpolation of the fluid velocity at the Lagrangian point, $\mathbf{U}^e$ from eq. \ref{rksubstep:b}, and the evaluation of the IBM force at the Lagrangian point, $\mathbf{F}(\mathbf{X}_{\eta})$ from eq. \ref{rksubstep:c}.  
    The next operation is the spreading operation (eq. \ref{rksubstep:d}), where the computed Lagrangian force is distributed into the adjacent Eulerian grid cells, as sketched in fig. \ref{fig:force_lag}b.  
    Note that this is a global operation, which requires all threads to add their contributions to a buffer in the {\it global} memory. 
    Since the execution of the threads is asynchronous, this operation can produce race conditions. 
    We avoid this problem using  {\it atomic} operations \cite{atomic} to orderly add each thread contribution, ensuring that each thread can read, modify, and write values to the {\it global} memory without interference from other threads. 

    \item \textbf{L2}, {\it Lagrangian Update}: This kernel updates the positions ($\mathbf{X}_{\eta}$) and surface velocities ($\mathbf{U}^{d}_{\eta}$) of the Lagrangian points discretizing the surface of the body. 
    Each thread is tasked with computing the data for a specific Lagrangian point. 
    The details of this kernel is strongly dependent on the coupling between the body deformation/dynamics and the fluid. 
    For problems with prescribed motion, like the ones considered here, synchronization between the threads is not required for read/write operations since each thread handles a distinct output.
    \end{itemize}

An alternative approach could consist of splitting the current calculation of the IBM forcing term (L1 kernel) into two different kernels: the first one with a one-dimensional grid architecture, in charge of the Eulerian-Lagrangian interpolation and subsequent calculation of $\mathbf F(\mathbf X_\eta)$; and the second one computing the Lagrangian-Eulerian spreading, employing a three-dimensional grid architecture as proposed in kernels E1, E2 and E3. 
This approach would avoid the utilization of atomic operations, since each thread, associated to a single Eulerian cell, would be responsible of gathering and adding the different portions of $\mathbf F$ associated to all the neighbouring Lagrangian points within their scope.
However, it would not benefit neither from the load balancing in the computation of $\mathbf f(\mathbf x_\xi)$, nor from the reuse of the evaluation of delta functions proposed for the single-kernel approach (L1).
Additionally, it would likely imply a slight overhead related to the initialization of a second kernel with a grid architecture different to the previous one, which in the single-kernel approach adopted in L1 is non-existent.
All these observations, together with the fact that a similar strategy has been employed in other works \cite{wu2019gpu}, have motivated the choice of a single kernel configuration as the one proposed in L1.

\subsubsection{Solution of systems of linear equations}
\label{sec:LinSystems}

Regarding the Helmholtz problem outlined in equations \eqref{rksubstep:e} and the Poisson equation \eqref{rksubstep:f}, 
we have chosen a direct method based on Fourier transformations (FFTs). 
The rationale behind this choice is two fold. First, the FFT subroutines of the CuFFT library \cite{cuda} provided by CuPy show excellent performance. 
Second, the single-GPU implementation implies that the matrix transpositions required to perform FFTs in three directions are done in the {\it global} memory, with negligible overhead. 
%
%The drawbacks of this choice is the limitation in the boundary conditions of the computational domain, \JMCr{although in practice a wide range of problems can be tackled with this approach.} 
% 
%Hence, 
To provide some flexibility with the boundary conditions, 
we have developed two different branches of TUCANGPU. The first one (TUCANGPU-3P) features a triply-periodic computational domain, and uses FFTs in three directions to diagonalize the systems of linear equations in eqs.  \eqref{rksubstep:e} and \eqref{rksubstep:f}. 
The second one is TUCANGPU-2P, with a computational domain with periodic boundary conditions in two dimensions (namely $y$ and $z$) and non-periodic boundary conditions along the $x$ direction.
In this branch, FFTs are used along the periodic directions to reduce the systems of $N = N_x\times N_y \times N_z$ linear equations into $N_y \times (N_z/2 +1)$ systems of $N_x$ complex linear equations, featuring tri-diagonal matrices, with $N_x,N_y,N_z$ defined as the number of cells per direction $x,y$ and $z$.
In both cases, the finite difference approximations to the spatial derivatives are realized in Fourier space using the concept of the {\it modified wave number} $\kappa'$, which can be obtained by applying the corresponding finite difference formula to a Fourier mode with wavenumber $\kappa$ (see \cite{moin2010fundamentals} and example in appendix \ref{sec:wavenumber}).
The subroutines developed for the triply-periodic computational domain, TUCANGPU-3P, are: 
\begin{itemize} 
\item{\bf S1}, \textit{Helmholtz problem}. This subroutine computes $\mathbf u^*=(u^*,v^*,w^*)$, solving eq. \eqref{rksubstep:e} once the corresponding right-hand-side has been computed by E1 and L1.  
Note that the Helmholtz problem for each velocity component can be solved independently. 
Applying the 3D FFT to the velocity component $u^*$ we obtain: 
\begin{equation} 
\nabla^2 u^* - \frac{\mathrm{Re}}{\beta_n\Delta t} u^* = \mathrm{RHS}_u 
\hspace{5mm} 
\xrightarrow{\displaystyle{\mathscr{F}_{3D}}}
\hspace{5mm} 
- \left( | \boldsymbol{\kappa}' |^2  + \frac{\mathrm{Re}}{\beta_n\Delta t} \right) \widehat{u^*} = \widehat{\mathrm{RHS}_u},
\end{equation}
were the hat ( $\hat{}$ ) superindex is used to denote variables in Fourier space, $\mathscr{F}$ denotes Fourier transformation,  and $| \boldsymbol{\kappa}'|^2$ contains the modified squared-wavenumbers that correspond to a second order, centered, finite difference approximation to the Laplacian operator (see appendix \ref{sec:wavenumber}).  
Hence, subroutine S1 first computes the 3D real FFT of $\mathrm{RHS}_u$, then solves for $\widehat{u^*}$, and perform an inverse FFT (iFFT) to compute $u^*$. This procedure is applied to each velocity component of $\mathbf{u}^*$. 

\item{\bf S2}, \textit{Poisson problem}. This subroutine computes $\phi$, solving eq. \eqref{rksubstep:f} once the corresponding right-hand-side has been computed by the Eulerian kernel E2. The procedure is analogous to S1, without the extra term of the Helmholtz problem. 
Hence, subroutine S2 first computes the 3D real FFT of $\mathrm{RHS}_\phi$, then solves for $\widehat{\phi} = -\widehat{\mathrm{RHS}_\phi}/ | \boldsymbol{\kappa}' |^2$, and performs an iFFT to compute $\phi$. 
It is worth noting that in this case the Poisson problem's system of equations is underdetermined for the first Fourier mode, $\kappa'_x=\kappa'_y=\kappa'_z=0$, indicating an infinite number of potential solutions. 
To tackle this, a common strategy involves imposing constraints on the solution. 
In this context, we opt to enforce the mean value of the pseudo-pressure to zero for all time instants $t$, expressed as $\widehat{\phi}(0,0,0,t)= 0$.

\end{itemize} 

For the 2D periodic branch, TUCANGPU-2P, the subroutines required to solve the linear systems of equations are: 
\begin{itemize} 
\item{\bf S1.1}, LU solution for the Helmholtz problem. The solution of eq. \eqref{rksubstep:e} in the TUCANGPU-2P is performed after a 2D FFT: 
\begin{equation} 
\nabla^2 u^* - \frac{\mathrm{Re}}{\beta_n\Delta t} u^* = \mathrm{RHS}_u 
\hspace{5mm} 
%\rightarrow^{FFT^2D} 
\xrightarrow{\displaystyle{\mathscr{F}_{2D}}}
\hspace{5mm} 
 \frac{\mathrm d ^2 \,\widehat{u^*}_{jk}}{\mathrm d x^2} - \left( | \boldsymbol{\kappa}_{jk}' |^2  + \frac{\mathrm{Re}}{\beta_n\Delta t} \right) \widehat{u^*}_{jk} = \widehat{\mathrm{RHS}_{u,jk}},
\end{equation}
where $\widehat{u^*}_{jk}$ is a vector of $N_x$ elements, with the 2D Fourier coefficients corresponding to the $j$-th wavenumber in $y$ and the $k$-th wavenumber in $z$, and $| \boldsymbol{\kappa}_{jk}' |^2$ is the modified wavenumber corresponding to second-order finite differences.
Discretizing the spatial derivative $\mathrm d^2/\mathrm dx^2$ with second order finite differences results in a tri-diagonal system of equations for $\widehat{u^*_{jk}}$, that reads:

\begin{equation}
\label{eq:tridiagsys}
    \frac{1}{h^2} \widehat{u^*}_{i-1,jk}- \left(\frac{2}{h^2}+ | \boldsymbol{\kappa}_{jk}' |^2 + \frac{\text{Re}}{\Delta t \beta_n} \right)\widehat{ u^*}_{i,jk}+\frac{1}{h^2} \widehat{u^*}_{i+1,jk}  =  \widehat{\mathrm{RHS}_{u}}_{,i,jk} ,\\
\end{equation}

\begin{figure}[t!]
    \centering
    \includegraphics[width=0.95\textwidth]{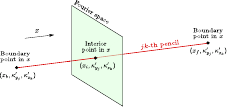}
    \caption{Schematic figure of the $jk$-th pencil along the non-periodic direction $x$. The first grid point (boundary) in this direction is denoted by $x_b$, the final boundary point by $x_f$, and a generic internal point is represented by $x_i$. This $jk$-th pencil containts the distribution of the Fourier modes $\kappa'_y,\kappa'_z$ along the non-periodic direction. Boundary conditions are set at the boundary points $(x_b,x_f)$, using their respective Fourier modes.}
    \label{fig:pencils}
\end{figure}

Again, the system of equations for each 2D Fourier mode is completely decoupled from the rest of modes, and thus can be solved independently, using a \textit{pencil} for each Fourier mode $jk$ as depicted in fig. \ref{fig:pencils}.
This allows us to solve them in parallel.
The pencil-like approach proposed herein has been similarly used in previous works \cite{zhu2018afid,van2015pencil,vela2021low} in the field of wall-bounded turbulence. In these studies, pencils are employed along the wall-normal direction. Refer to appendix \ref{sec:appendix_linearsystems} for more details.

Consequently, the kernel S1.1 uses a GPU grid decomposition based on the 2D FFT, where each thread solves the tri-diagonal system in eq. \eqref{eq:tridiagsys} for the $jk$-th Fourier mode.
The kernel is fed with the 2D real FFT of $\mathrm{RHS}_{u^*}$, 
and the solution of the tri-diagonal system is performed with the Thomas' algorithm, which provides a low-storage LU factorization.
The output of each thread is $\widehat{u^*}_{jk}$, which is written in the {\it global} memory of the GPU card. 
Note that this kernel is called three times, one for each velocity component. 

\item{\bf S1.2}, iFFT to compute $\mathbf{u}^*$. Since the output of S1.1 is in Fourier space, it is necessary to perform a 2D iFFT of $\widehat{u^*}_{jk}$ to obtain $\mathbf{u}^*$. This is a global operation in the $y$-$z$ planes, and needs to be performed once the kernel S1.1 has finished. 

\item{\bf S2.1}, LU solution for the Poisson problem. As before, the solution of eq. \eqref{rksubstep:f} in TUCANGPU-2P is performed by feeding the kernel with the 2D real FFT of $\mathrm{RHS}_{\phi}$. Analogously, the solution of the tri-diagonal systems is performed with the same LU factorization employed before.
As a matter of fact, after discretizing the second-order derivatives in $x$, the coefficient multiplying the diagonal element is the same as in the Helmholz problem except for the term containing the Reynolds number.
Similarly to the periodic case, when using Neumann conditions at both boundaries, the first Fourier mode $\kappa'_y=\kappa'_z=0$ (the mean value of the pseudo-pressure) becomes underdetermined. 
Consequently, a value must be assigned at a specific $x$-position for this mode, to remove the singularity.
We have set this value equal to zero at the first interior point of the pseudo-pressure, i.e. $\widehat{\phi}(x_b+h/2,0,0,t) = 0$, where $x_b$ is the $x$-coordinate of the first physical boundary in the non-periodic direction $x$.
See appendix \ref{sec:appendix_linearsystems} for additional details.

\item{\bf S2.2}, iFFT to compute $\phi$. Since the output of S2.1 is in Fourier space, it is necessary to perform a 2D iFFT of $\widehat{\phi}_{jk}$ to obtain $\phi$. This is a global operation in the $y$-$z$ planes, and needs to be performed once the kernel S2.1 is finished.

\end{itemize}

\subsubsection{Subroutines to set up boundary conditions}

In order to impose boundary conditions in the evaluation of the explicit terms appearing in eq. \eqref{rksubstep:main}, TUCANGPU uses ghost cells in the boundaries of the computational domain \cite{hirsch2007numerical}. 
The subroutines in this group copy boundary data to the corresponding ghost cells to impose Dirichlet, Neumman or periodic boundary conditions, 
enabling the use of the same stencil for the finite differences operators of the Eulerian kernels for internal and boundary points. 

Subroutine BC1 sets the boundary conditions for $\mathbf{u}$ and $p$ in the periodic directions. 
For TUCANGPU-2P, subroutine BC2.1 sets the boundary conditions along the non-periodic direction $x$, which are stored in the ghost cells of the buffers for $\mathbf{u}$ and $p$. 
Note that in TUCANGPU-2P, we also need boundary conditions for the 1D Helmholtz and Poisson problems obtained in section \ref{sec:LinSystems} for each Fourier mode. 
These are computed in subroutine BC2.2, performing the 2D FFTs of the corresponding ghost cell planes using the CuFFT library. 
All operations in BC1, BC2.1 are performed directly in the {\it global} memory of the GPU-card, and are coded in TUCANGPU using CuPy.

%%%%%%%%%%%%%%%%%%%%%%%%%%%%%%%%%%%%%%%%%%%%%%%%%%%%%%%%%%%%%
%%%%%%%%%%%%%%%%% Verification
%%%%%%%%%%%%%%%%%%%%%%%%%%%%%%%%%%%%%%%%%%%%%%%%%%%%%%%%%%%%%
\section{Verification} \label{sec:verification}
To verify the results obtained with TUCANGPU, we have performed a comparison with TUCAN \cite{moriche2017numerical,gonzalo2018aerodynamic}, the CPU version of TUCANGPU. 
The test problems chosen for the verification of TUCANGPU are the flow around a fixed and a moving sphere as baseline cases.
For the case of a moving sphere, we prescribe a sinusoidal motion along the $y$-direction with expression $y(t)/D = \sin(\frac{2\pi}{5} t)$, where $D$ is the diameter of the sphere. 
Each case is solved using either a three-periodic domain (TUCANGPU-3P) or an inflow-outflow configuration with lateral periodic boundaries (TUCANGPU-2P). This totals four cases, all computed in double precision for both CPU and GPU simulations.

The domain size is $10.24D \times 5.12D \times 5.12D$ for all the cases and the grid spacing corresponds to 25 points per diameter ($D/\Delta x = 25$). 
The Reynolds number based on the sphere diameter $D$ and the initial free-stream velocity $U_{\infty}$ is $\mathrm{Re}_D=U_\infty D/\nu =200$, where $\nu$ is the kinematic viscosity.
The sphere is centered at the origin, at a distance of $3D$ from the inlet in the $x$-direction for the inflow-outflow configuration (note that this distance is not relevant in the triply-periodic setup, since there is no inlet).
All the simulations have been run with a fixed time-step, such that the Courant-Friedrichs-Lewy number (CFL) is around $0.3$.
The initial condition for the simulations is set as a uniform flow in the $x$-direction, represented by the velocity vector $\mathbf{u} = (U_{\infty}, 0, 0)$.

%%%%%%%%%%%%%%%%%%%%%%%%%% Full-periodic
\subsection{Triply-Periodic}\label{sec:fullperiodic}
\begin{figure}[t!]
    \centering
    \hspace*{-1cm}
    \includegraphics[width=1.15\textwidth]{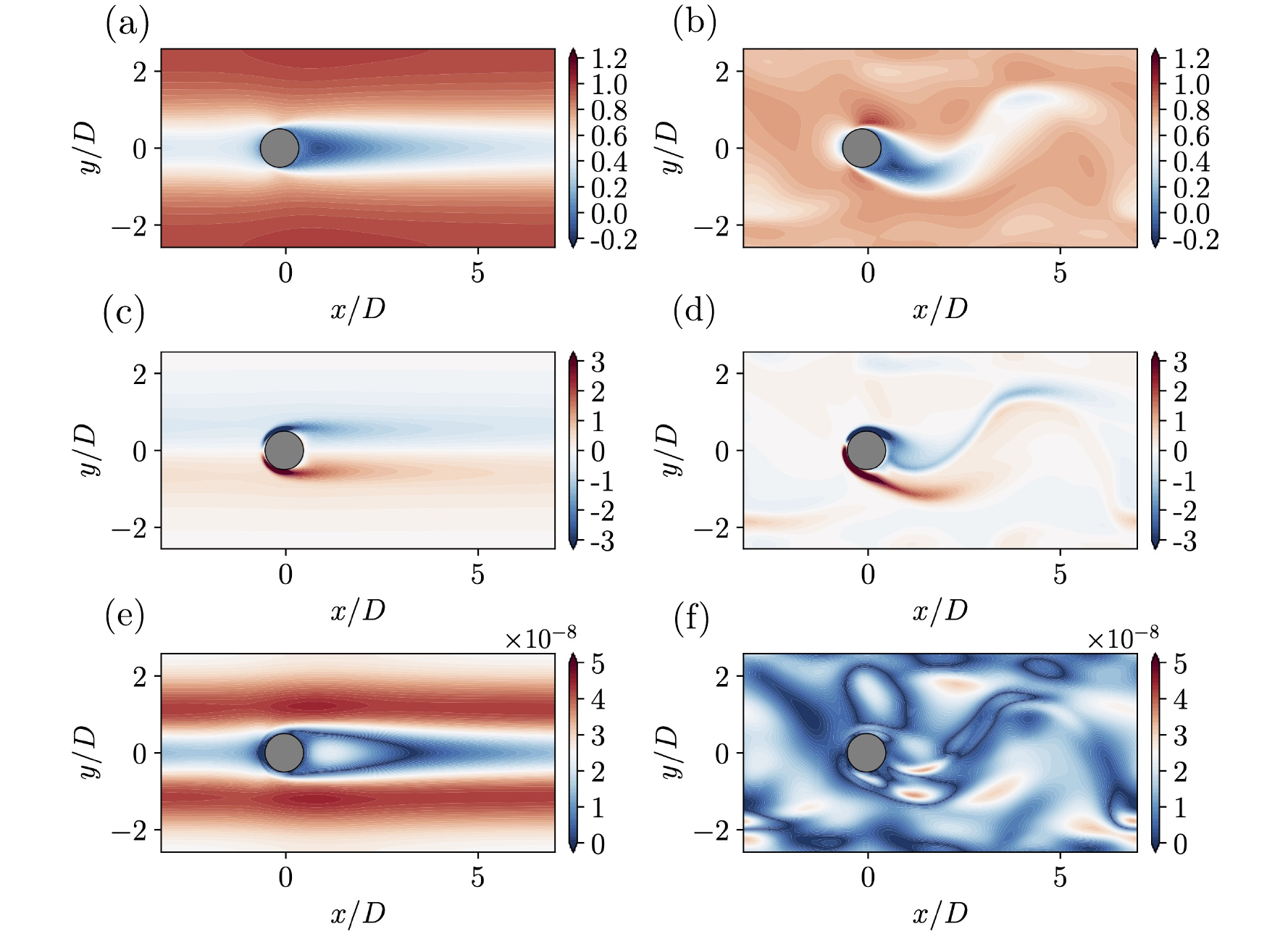}
    \caption{Periodic flow around sphere at $z/D = 0$ at $tU_\infty/D = 140$. Left column corresponds to a fixed sphere, right column to a moving sphere. From top to bottom: streamwise component of the velocity, $u/U_\infty$, $z$-component of the vorticity, $\omega_z D /U_\infty$, and instantaneous $u$ absolute error after $140$ convective time units between TUCAN and TUCANGPU-3P.}
    \label{fig:periodicflow}
\end{figure}

Focusing first on the results of the fixed sphere, figs. \ref{fig:periodicflow}a,c show instantaneous snapshots of streamwise velocity $u$ and out-of-plane vorticity $\omega_z$ at $z/D = 0$, at $tU_\infty/D = 140$. 
As seen, the simulation reaches a symmetric solution, characterized by a symmetric wake after $140$ convective time units.
For what regards the case with the moving sphere, figs. \ref{fig:periodicflow}b,d show again instantaneous results of $u$ and $\omega_z$ at $tU_\infty/D = 140$, demonstrating periodic shedding dominated by the frequency of the body's motion.

To provide a visual representation of the discrepancies between the CPU and GPU versions, figs. \ref{fig:periodicflow}e,f present the absolute difference between the results obtained by the two solvers on the plane $z/D = 0$ for both kinematics. As seen, the values of these differences are $\mathcal{O}(10^{-8})$ everywhere.
Quantifying these differences further, table \ref{tab:max_error} displays the maximum absolute errors observed across the entire domain for the velocity components $u$, $v$ and $w$, as well as for the pressure $p$. 
The magnitude of absolute error between TUCAN and TUCANGPU-3P is $\mathcal{O}(10^{-8})$ after $140$ convective times, verifying the implementation the GPU-accelerated version.
This error is consistent with the tolerance level of the iterative solver, $\mathcal{O}(10^{-9})$ in TUCAN CPU, given that GPU version solves these systems directly using FFT transformation.
\begin{table}
\centering
\begin{tabular}{>{\centering}p{0.25\linewidth}
                >{\centering}p{0.12\linewidth}
                >{\centering}p{0.08\linewidth}
                >{\centering}p{0.08\linewidth}
                >{\centering}p{0.08\linewidth}
                >{\centering}p{0.08\linewidth}
                >{\centering\arraybackslash}p{0.08\linewidth}}
\multicolumn{7}{c}{} \\
   Streamwise boundary conditions & Sphere kinematics & $\varepsilon_x$ & $\varepsilon_y$ & $\varepsilon_z$ & $\varepsilon_p$ &   \\\midrule
Periodic & Fixed & $4.44$     & $1.64$     & $1.64$ & $1.75$ & $\times 10^{-8}$ \\
Periodic & Moving & $5.19$     & $4.33$     & $2.26$ & $2.82$ & $\times 10^{-8}$ \\
Inflow-outflow& Fixed & $3.87$     & $3.26$     & $3.24$ & $5.86$ & $\times 10^{-8}$ \\
Inflow-outflow& Moving & $7.91$     & $4.81$     & $3.28$ & $3.13$ & $\times 10^{-8}$ \\ \bottomrule
\end{tabular}
\caption{Maximum absolute errors for $u_x$,$u_y$,$u_z$ and $p$ for both configurations}
\label{tab:max_error}
\end{table}

%%%%%%%%%%%%%%%%%%%%%%%%%% Inflow-outflow
\subsection{Inflow-Outflow}\label{sec:inflow-outflow}
\begin{figure}[t!]
    \centering
    \hspace*{-1cm}
    \includegraphics[width=1.15\textwidth]{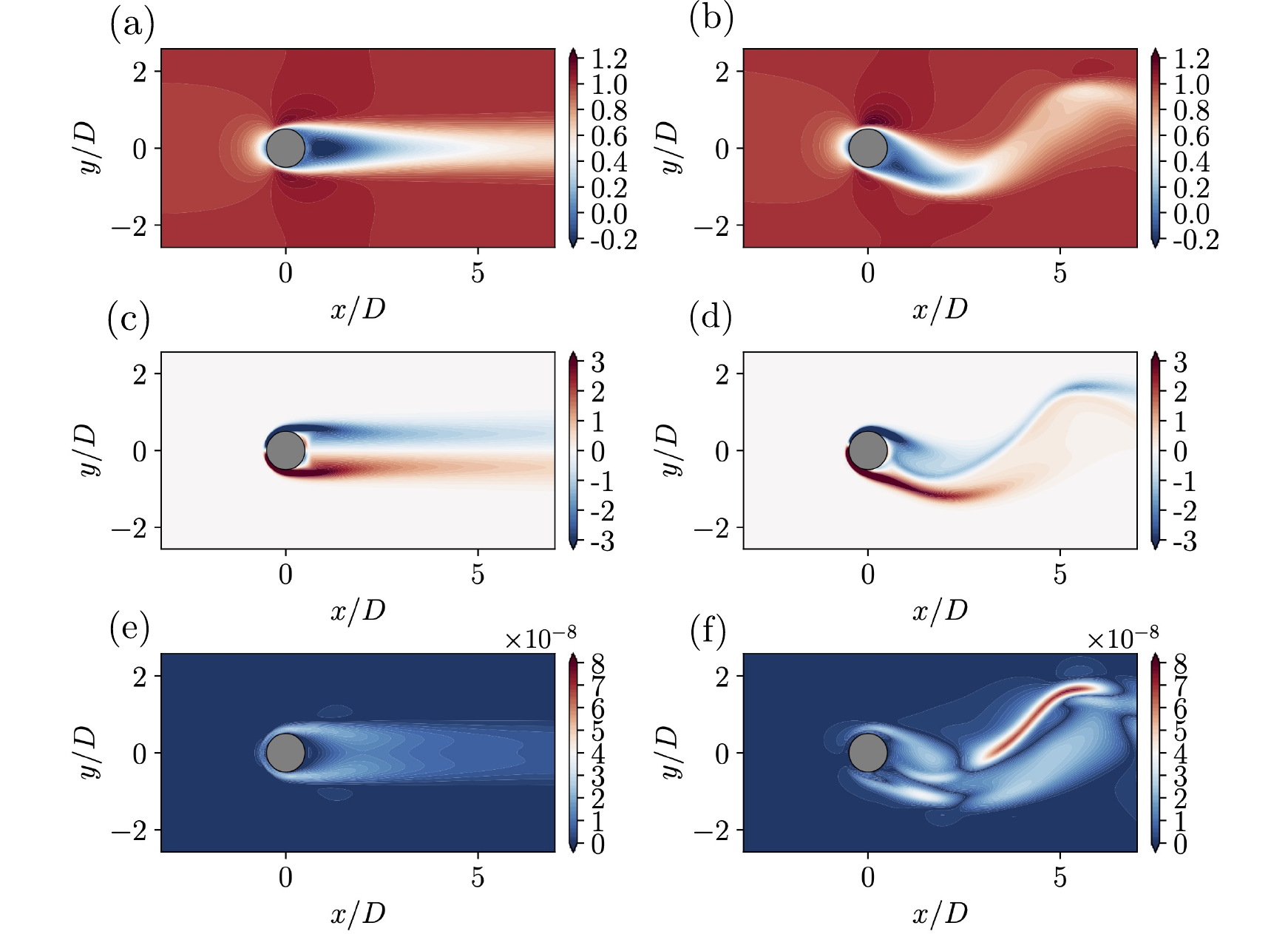}
    \caption{Flow around sphere at $z/D = 0$ at $tU_\infty/D = 140$ in an inflow-outflow streamwise configuration. Left column corresponds to a fixed sphere, right column to a moving sphere. From top to bottom: streamwise component of the velocity, $u/U_\infty$, $z$-component of the vorticity, $\omega_z D /U_\infty$, and instantaneous $u$ absolute error after $140$ convective time units between TUCAN and TUCANGPU-2P.}
    \label{fig:in-out}
\end{figure}

We center now our attention to the inflow-outflow configuration.
Dirichlet boundary conditions are set at the inlet ($x=x_b$) as $\mathbf u(x_b,y,z,t) = (U_{\infty},0,0,t)$ for any simulation time $t$. At the outlet ($x=x_f$), an advective boundary condition is applied:

\begin{equation}
    \frac{\partial \mathbf u}{\partial t} + C\frac{\partial \mathbf u}{\partial x} = 0,
\end{equation}
where $C$ represents the advection velocity, assumed to be constant along the outlet. More details on the implementation of this boundary condition can be found at \cite{moriche2017}.
For the pressure and pseudo-pressure we impose homogeneous Neumann boundary conditions at both inlet and outlet, expressed for the pressure as $\frac{\partial p}{\partial x}(x_b,y,z,t) = \frac{\partial p}{\partial x}(x_f,y,z,t) = 0$.
Figures \ref{fig:in-out} shows instantaneous snapshots of streamwise velocity $u$ (figs. \ref{fig:in-out}a,b) and vorticity $\omega_z$ (figs. \ref{fig:in-out}c,d), for a fixed and a moving sphere, respectively, in the inlet/outlet configuration. 
Similarly to the triply-periodic case,  figs. \ref{fig:in-out}a,c exhibit a steady symmetric solution, even after 140 convective time units, whereas figs. \ref{fig:in-out}b,d present again a wake that is tilted and changes according to the prescribed motion of the sphere. 

With the objective of highlighting the differences between the results obtained the with the two solvers, figs. \ref{fig:in-out}e,f depict the absolute difference between them for the streamwise velocity field $u/U_{\infty}$ at $z/D=0$. 

Once more, we observe that the measured errors consistently fall within $\mathcal{O}(10^{-8})$ throughout the domain (see also table \ref{tab:max_error}). This includes the maximum errors across all Eulerian variables, not limited to just the streamwise velocity, which remain at this same order even after $140$ convective time units. Similar to the triply-periodic case, these discrepancies align with the tolerance of the iterative solver employed in TUCAN.

%%%%%%%%%%%%%%%%%%%%%%%%%%%%%%%%%%%%%%%%%%%%%%%%%%%%%%%%%%%%%
%%%%%%%%%%%%%%%%% Code performance
%%%%%%%%%%%%%%%%%%%%%%%%%%%%%%%%%%%%%%%%%%%%%%%%%%%%%%%%%%%%%
\section{Code performance} \label{sec:code-performance}

In this section we assess the performance of TUCANGPU as compared to its analogous CPU version, TUCAN.
We consider the flow around a fixed sphere, on the same domain, Re and CFL as in section \ref{sec:verification}.
In this test, two different grid sizes are considered with dimensions $256 \times 128 \times 128$ and $512 \times 256 \times 256$. Hereafter, these two cases will be referred as low and high resolution, respectively.

GPU experiments were executed using a TITAN V NVIDIA GPU equipped with 12 GB of RAM, 5120 CUDA cores, and 640 tensor cores of first generation. 
Conversely, CPU tests were conducted on a mid-sized cluster comprising 22 heterogeneous nodes, featuring Intel Xeon X5650, E5-2630 and E5-2620v4 processors, interconnected through a dedicated InfiniBand QDR network.

\begin{figure}
    \centering
    \includegraphics[width= 1\textwidth]{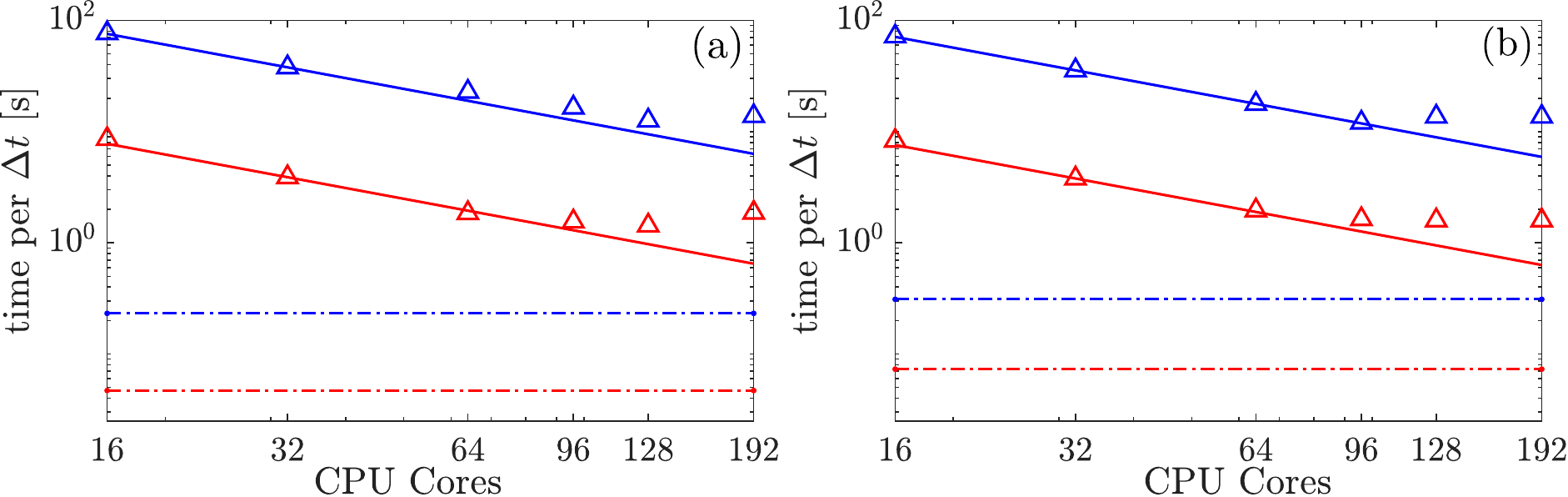}
    \caption{Wall-time per $\Delta t$ for the triply-periodic case (a) and the inflow-outflow case (b). Legend:  GPU (dashed), CPU (markers). The solid lines describe the ideal iteration time scaling with the number of processors. Red color represents low resolution, while blue color refers to high resolution.}
    \label{fig:perfo_periodic_inout}
\end{figure}

Figure \ref{fig:perfo_periodic_inout} shows the wall-time per $\Delta t$ of TUCAN and TUCANGPU for the two grid-sizes and flow configurations considered.
The results demonstrate that for the high-resolution case under triply-periodic and inlet-outlet conditions, weak scaling for the CPU code remains linear up to 128 and 96 processors, respectively.
The optimal iteration time in the triply-periodic case is achieved with 128 cores, taking $1.43$s and $12.57$s for low and high resolutions, respectively.
On the other hand, inlet-outlet case achieves its optimal with 96 CPU cores with $1.6$s and  $12.04$s for low and high resolution, respectively.
A comparison between resolutions in both configurations tests the strong scaling of the solver, demonstrating that the wall-time per time step is proportional to the domain size. Specifically, an 8x increase in the number of points results in an 8x increase in time, indicating that the solver exhibits effective scalability.
However, the CPU parallel efficiency is compromised well before reaching GPU speeds.
Note that the cluster used for the tests with the CPU version of TUCAN  has 16 cores per node. 
Hence, the first data point in figs. \ref{fig:perfo_periodic_inout}a,b corresponds to a case in which no inter-node communications take place.

Comparatively, in the triply-periodic cases the GPU version is 30.5 and 54.2 times faster than the CPU with peak performance, for low and high resolutions respectively.
In the inlet-outlet cases, alternatively, the GPU version surpasses CPU performance by 21.8 times for low and 38.8 times for high resolutions.
In other words, assuming ideal scalability in TUCAN, the number of CPUs needed to match GPU performance should be around 2500 for low resolution and 5500 for high resolution.

\begin{figure}
    \centering
     \includegraphics[width= \textwidth]{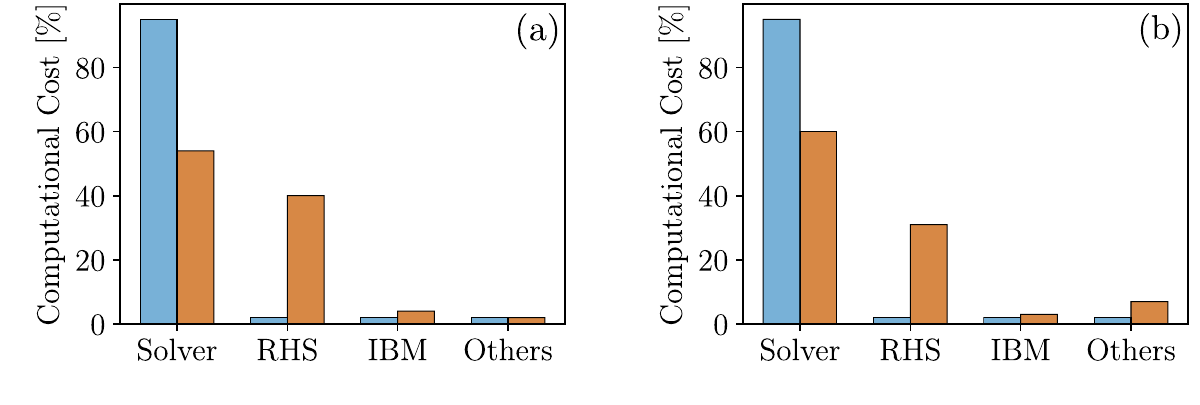}
    \caption{Bar plots showing the relative contribution to the total computational cost, per time step, of the different tasks performed for the triply-periodic (left) and the inflow-outflow configurations (right), in CPU (\protect\bluesquare) and GPU (\protect\orangesquare). }
    \label{fig:barplot_perfo}
\end{figure}

Figure \ref{fig:barplot_perfo} compares the distribution of computational cost of the tasks that both, CPU and GPU codes are required to execute, per time step $\Delta t$ for both boundary conditions setups: (a) tryply-periodic and (b) inflow-outflow.
These tasks comprise solving the Helmholtz and Poisson equations, computing explicit terms and performing IBM interpolations, denoted in the figures as Solver, RHS, and IBM respectively.
Regardless the boundary conditions, the CPU code requires 94\% of the computational resources for solving the linear systems, as seen in fig. \ref{fig:barplot_perfo}, from which most of the time is spent in the iterative solution of the Poisson problem.

In contrast, the GPU version significantly reduces this demand through directly solving the Poisson and Helmholtz problems through FFT transforms.
For TUCANGPU-3P, linear systems calculations account for only 53.5\% of the computational load, while TUCANGPU-2P allocates 59.7\% of the total time to this task. The Fourier transform reduces the solver wall-time in TUCANGPU-3P by a factor of 50 to 80, with the speed-up increasing at higher resolution. In TUCANGPU-2P, where the $x$-direction is solved directly via LU matrix inversion, the speed-up ranges from 35 to 60 for low and high resolutions, respectively.
These results highlight the excellent performance of the CuFFT library in solving the Poisson and Helmholtz problems.

In this context, the computational time spent on solving the linear system becomes comparable to that of computing the explicit terms on the right-hand sides (RHS), constituting 39.44\% and 30.67\% for triple and double periodic systems, respectively. The speed-up in RHS, compared to TUCAN, ranges from 1.6 to 3, for the different resolutions. These computations include non-linear and viscous terms, intermediate velocity divergence, and corrections of pressure and velocity, all of which are performed similarly in both CPU and GPU codes.

Furthermore, the GPU architecture maintains its efficiency when computing the IBM interpolations, consuming only 4.2\% and 2.86\% of the total computational time for TUCANGPU-3P and TUCANGPU-2P, respectively. 
This is likely a consequence of employing a Lagrangian grid centric arrangement for the parallelization of the IBM interpolations, evenly distributing the calculations across all GPU threads. 
Also, by fitting all computations onto a single GPU, communication between CPU processors is avoided, enhancing the performance further.
These optimizations lead to a speed-up that increases with resolution, achieving a range of 7.6 to 12 times compared to the CPU version for the IBM operations in this test.
As an alternative, an Eulerian-only approach was also tested, including Lagrangian computations, but resulted in a considerable performance decrease compared to having both mesh centric arrangements. 
This is primarily attributed to the uneven distribution of Lagrangian points in the domain. 
Uneven distributions require more operations on blocks responsible for larger quantities of Lagrangian points, causing other thread blocks without any Lagrangian points to wait.

As a final note, we remark that TUCAN is a massively parallel CPU-based solver, designed to run larger problems than those used here as benchmark. 
Thus, some of the performance metrics presented before for TUCAN may be slightly impacted on these grounds.

%%%%%%%%%%%%%%%%%%%%%%%%%%%%%%%%%%%%%%%%%%%%%%%%%%%%%%%%%%%%%
%%%%%%%%%%%%%%%%% Applications
%%%%%%%%%%%%%%%%%%%%%%%%%%%%%%%%%%%%%%%%%%%%%%%%%%%%%%%%%%%%%

\section{Applications} \label{sec:applications}
As a proof of concept, we present applications representative of external and internal flows, for which TUCANGPU is suitable:
\begin{enumerate}
    \item The flow around a rotating winged-seed. We study the unsteady aerodynamics and vortex shedding around a winged-seed rotating at constant angular velocity. The geometry of the seed is simplified as in \cite{Arranz2018}. For this case we employ an inflow-outflow configuration (using TUCANGPU-2P) as the one described in sections \ref{sec:verification} and \ref{sec:code-performance}.
    \item The flow inside a simplified model of the left ventricle. We consider a simplified geometry of a ventricle moving throughout the cardiac cycle, which is immersed in a triply-periodic domain (using TUCANGPU-3P).
\end{enumerate}

Both applications are somehow representative of the the wide range of CFD problems that the software can address, each leading to different computational performance.
Indeed, the rationale behind these choices is to evaluate the performance in two problems for which the ratio of total number of Lagrangian points to total number of Eulerian points is completely different.
In external aerodynamics, the characteristic volume of the immersed body is much smaller than the fluid domain. Conversely, cardiovascular flow simulations involve internal flows within structures such as vessels, atria, and ventricles. These structures result in body volumes comparable to the fluid domain, leading to larger Lagrangian meshes and thereby increasing the ratio between Lagrangian and Eulerian points.

For this purpose, we include performance results obtained using TUCANGPU on the previous GPU hardware (TITAN V), and we incorporate results obtained with a NVIDIA A100 model, a high-end GPU equipped with 80 GB of RAM, 6912 CUDA cores, and 432 third-generation Tensor Cores.

%%%%%%%%%%%%%%%%%%%%%%%%%% Samara
\subsection{Flow around a rotating winged-seed}

\begin{figure}[t]
    \centering
    \includegraphics[width=\textwidth]{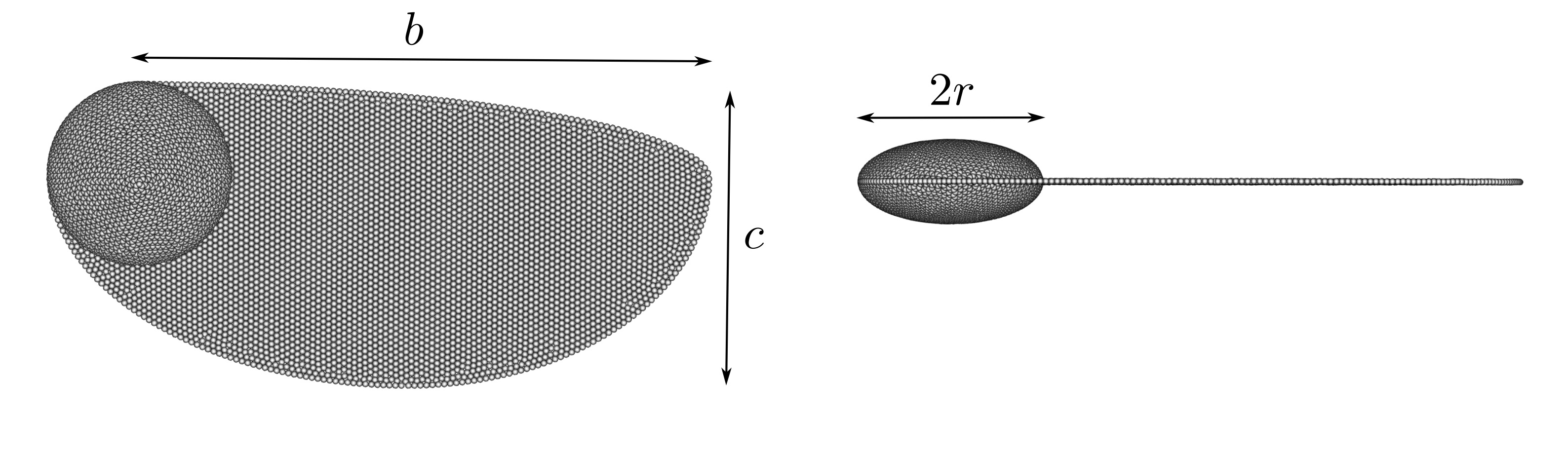}
    \caption{Top (left) and side (right) view of the Lagrangian grid for the winged-seed, showing the relevant lengths of the problem. Same configuration as \cite{Arranz2018}.}
    \label{fig:samara_scheme}
\end{figure}

We present in this section the flow past a rotating winged-seed. For this application, we choose a samara-type winged-seed, where its geometry is modeled using four tangent quarter of ellipses for the leaf \cite{pedersen2006indicial} and an oblate spheroid for the nut, as shown in fig. \ref{fig:samara_scheme}. 

\begin{figure}[t]
    \centering
    \includegraphics[width=\textwidth]{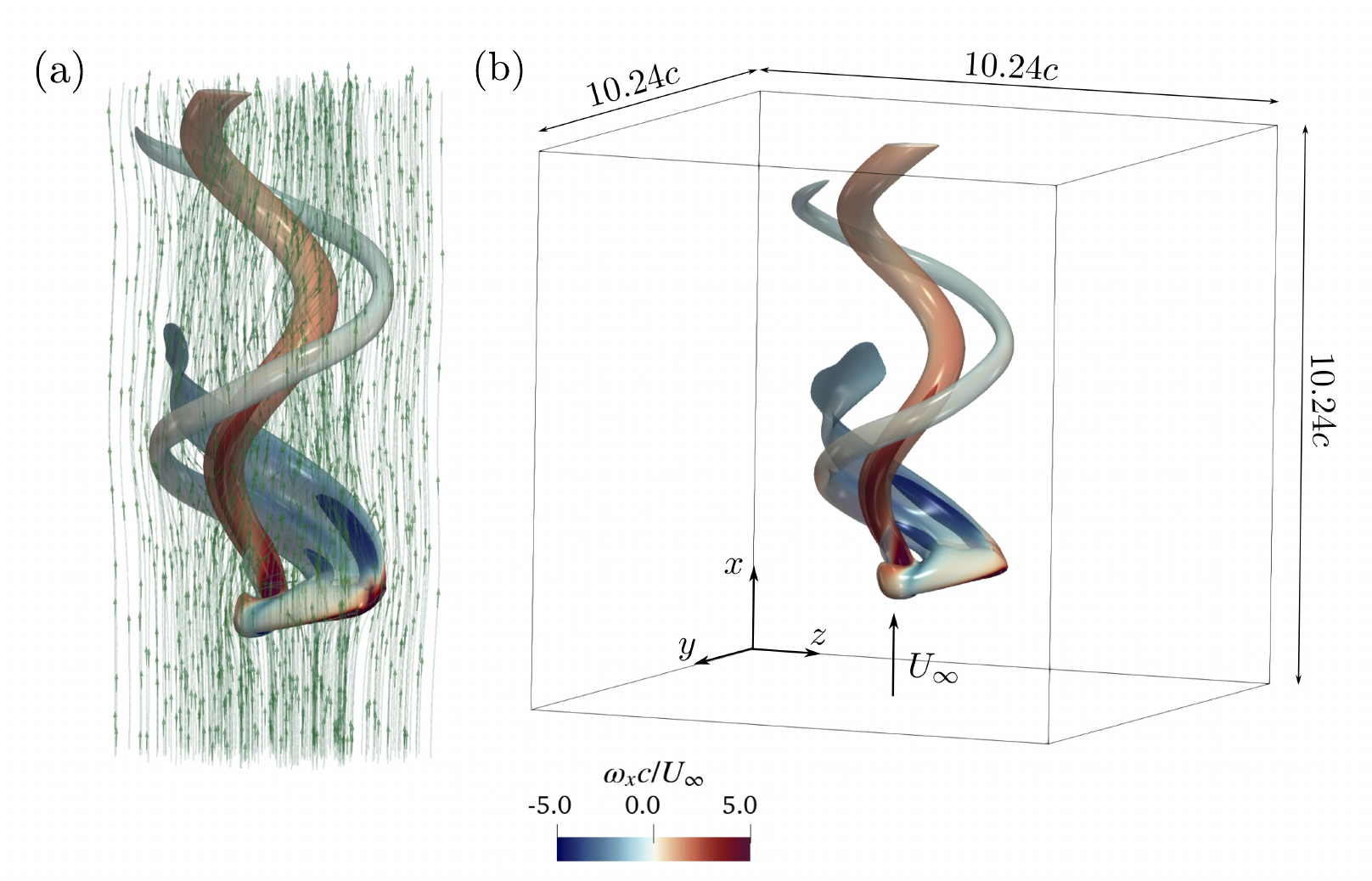}
    \caption{Instantaneous snapshots showing isosurfaces of nondimensional Q-criterion $Qc^2/U_{\infty}^2=[0.5,5]$, where the lower value is represented with a lower opacity. The isosurfaces are coloured with the non-dimensional $x$-component of the vorticity, $\omega_x c/U_{\infty}$. The vortical structures are shown together with streamlines and vectors pointing in the direction of the flow velocity (a), and the computational domain with its dimensions and the reference frame (b). }
    \label{fig:samara3Ds}
\end{figure}

\begin{figure}[t]
    \centering
    \hspace*{-0.5cm}
    \includegraphics[width=1.05\textwidth]{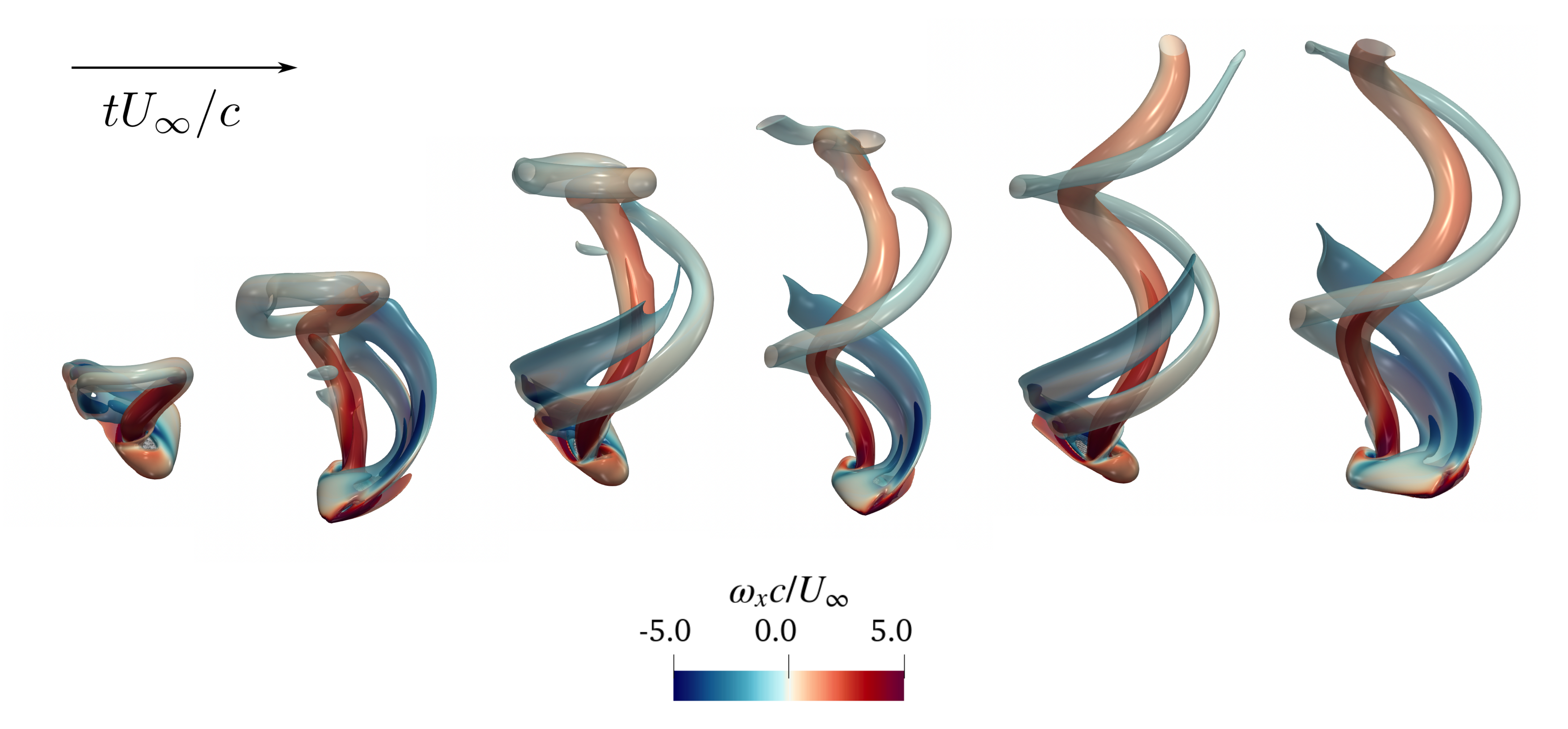}
    \caption{Vortical structures through the simulation transient. From left to right, different physical simulation times are shown, corresponding to $tU_{\infty}/c = [3,6,9,12,15,18]$. Same values of Q-criterion as in fig. \ref{fig:samara3Ds}.}
    \label{fig:samara3Ds2}
\end{figure}

\begin{figure}[t]
    \centering
    \hspace*{-0.5cm}
    \includegraphics[width=1.05\textwidth]{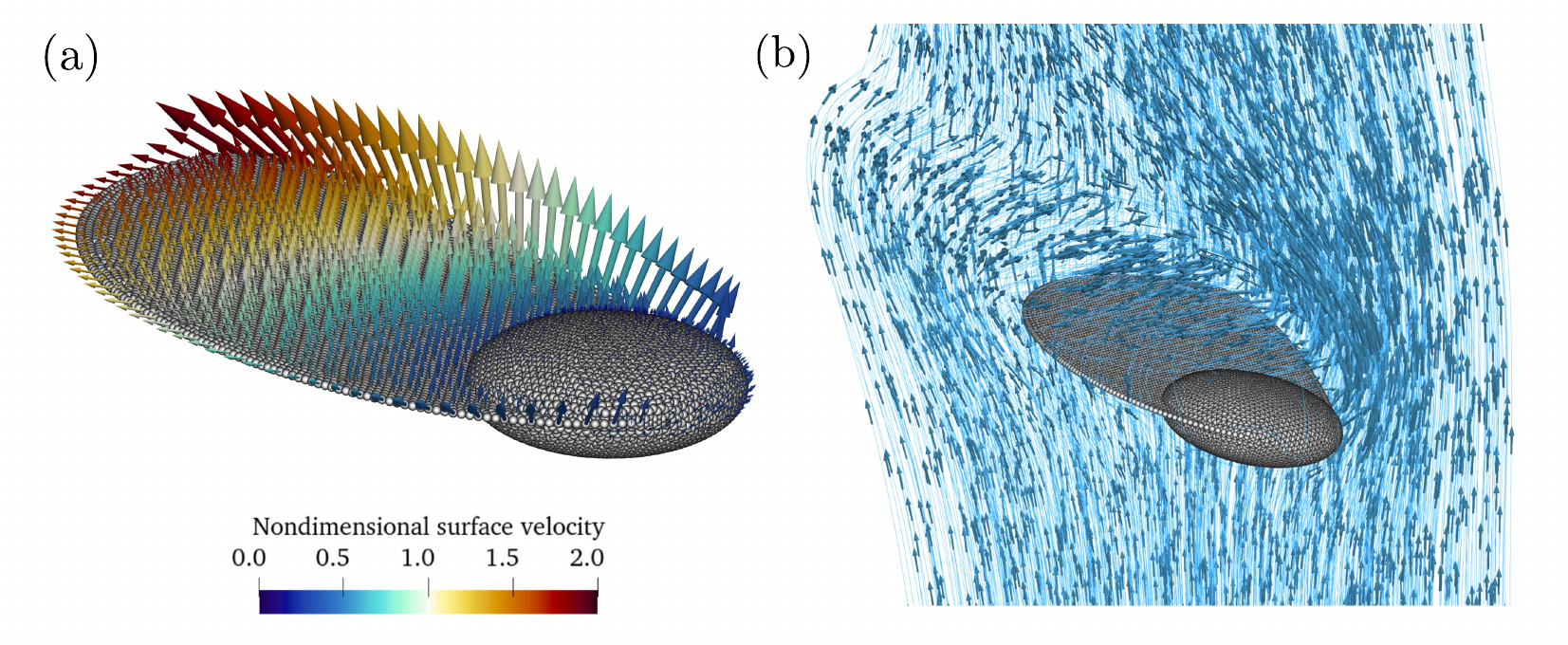}
    \caption{Three-dimensional visualizations of an instantaneous snapshot of the simulation, showing (a) a quiver plot with the distribution of aerodynamic forces on the surface of the winged-seed, scaled with the magnitude of the forces, and coloured by the magnitude of the non-dimensional surface velocity; and (b) vectors following the streamlines over the surface of the samara. For clarity, vectors in (a) are displayed every three Lagrangian points.}
    \label{fig:samara-quiver}
\end{figure}

Considering the chord of the wing $c$ as characteristic length of the problem, we can then define the remaining relevant lengths, such as the wing span $b/c=1.9$ and the semi-major axis of the nut $r/c=0.3$, whose aspect ratio is 0.6.

We choose inflow-outflow boundary conditions in the vertical direction $x$, and periodicity at the lateral directions $y,z$.
The Reynolds number based on the chord $c$, the inflow velocity $U_{\infty}$, and the kinematic viscosity $\nu$, is $\mathrm{Re}_c \equiv c U_{\infty}/\nu $ = 200. 
The domain size is $10.24c \times 10.24c \times 10.24c $ (see fig. \ref{fig:samara3Ds}b), where the number of grid points per chord is $c/\Delta x = 50$, yielding a domain with $512\times 512 \times 512$ grid points ($\sim 134$ millions).
To initialize the simulation, we select a uniform flow in the vertical direction $\mathbf u = (U_{\infty},0,0)$. Additionally, we place the samara at a distance of $2c$ from the inlet, and we select its attitude with a pitching angle $\theta = -15^{\circ}$ and a coning angle $\beta=10^\circ$, such that they are representative of a real falling winged-seed in auto-rotation (refer to \cite{Arranz2018} for more details on the definition of these angles).

During the simulation, the motion of the samara is governed by its rotation around a vertical axis through the center of the nut, with constant angular speed $\Omega = U_{\infty}/c$,
\begin{subequations}\label{eq:samara_kinematics}
\begin{align}
          x(t) & = x_0,\\
          y(t) & = y_0\cos(\Omega t) - z_0\sin(\Omega t),\\
          z(t) & = y_0\sin(\Omega t) + z_0\cos(\Omega t),
\end{align}
\end{subequations}
where $x_0,y_0$ and $z_0$ are the initial coordinates of the seed and $t$ is the physical simulation time. 

Figures \ref{fig:samara3Ds} and \ref{fig:samara3Ds2} show the vortical structures shed by the seed during its rotatory motion, exhibiting an helicoidal wake pattern consistent with its kinematics. 
This complex wake is roughly composed by two vortices of different sign. 
One of them is the vortex shed from the nut (in red), and the other is the one shed by the wing (in blue).
As the samara rotates, these two vortices, initially attached (see the higher Q-criterion value in figs. \ref{fig:samara3Ds} and \ref{fig:samara3Ds2}, with more opacity), start to detach from the surface of the wing, forming the aforementioned helicoidal vortical structure.

Focusing on fig. \ref{fig:samara3Ds2}, we observe that this process takes place progressively. The figure displays how the helicoidal wake forms and develops along the initial transient of the simulation.
We can inspect further this physical phenomena by taking a look to the distribution of aerodynamic forces over the samara (fig. \ref{fig:samara-quiver}a). 
In this figure, the size of the vectors is scaled with the magnitude of the aerodynamic forces, and most of them have a predominant vertical component, meaning that the fluid is exerting, on average, a vertical force over the rotating samara, in accordance with the direction of the free-stream. 
The vectors, additionally, are coloured by the magnitude of the surface velocity, that is increased as we approach the wingtip because of the rotation.
The magnitude of the forces is the strongest along the leading edge, due to the spanwise suction peak generated in the attached leading edge vortex (LEV). 
The aerodynamic load, however, decreases along the chordwise direction, as approaching the trailing edge, where the influence of the LEV is significantly weaker. 
Interestingly, we find that the aerodynamic load along the trailing edge has mainly a component roughly tangential to the wing, indicating that there is no flow separation in this part of the wing.
Contrary, in the vicinity of the nut, the flow is fully-detached, as seen in fig. \ref{fig:samara-quiver}b.
Note that this separation takes place because the vertical velocity $U_\infty$ is considerably higher than the rotational velocity in this region. 

\begin{figure}[h]
    \centering
    \includegraphics[width= 1\textwidth]{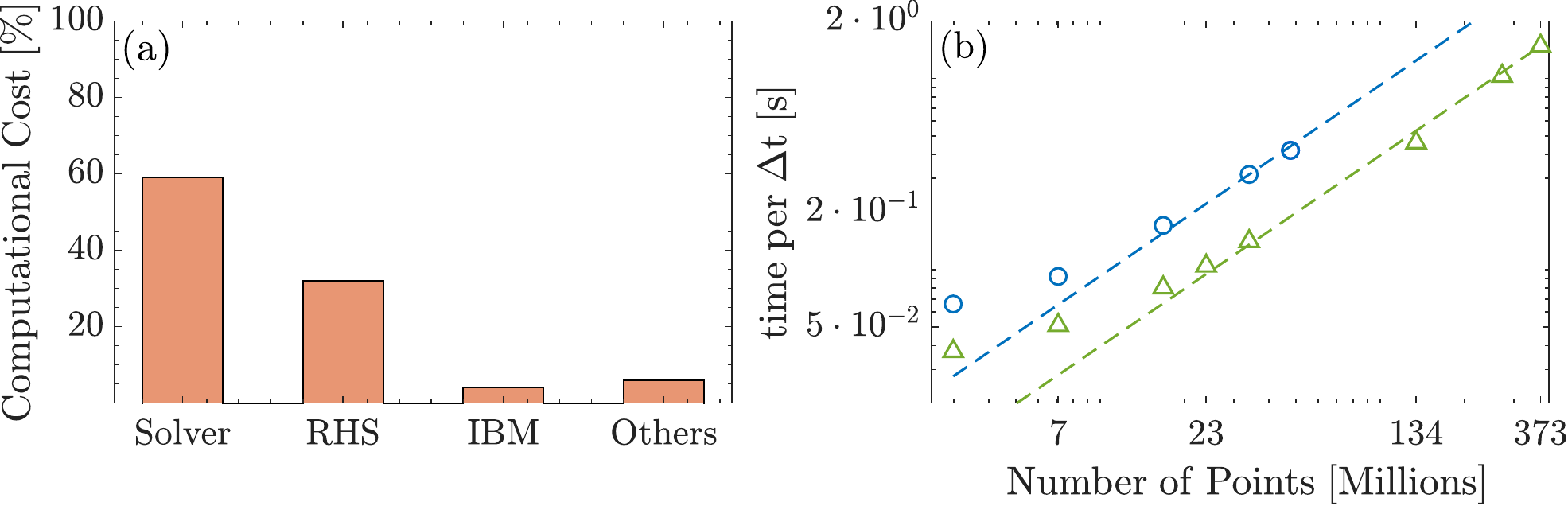}
    \caption{Performance indicators in the winged-seed problem. (a) Bar plots for the relative contribution of different tasks performed on the GPU to the total computational cost per time step. (b) The wall-time per $\Delta t$ is displayed for different domain resolutions. Dashed lines represent the ideal strong scaling with domain size for two different GPU hardware: TITAN V (\protect\dashedclearblueline) and A100 (\protect\dasheddarkergreenline).}
    \label{fig:samara_perfo}
\end{figure}

To evaluate the efficiency and scalability of TUCANGPU-2P, we assess its performance in double precision. %and single precision. 
In fig. \ref{fig:samara_perfo}a, the relative contribution of GPU-performed tasks to the overall computational cost per time step is illustrated, showing a comparable distribution of the relative computational costs among the Solver, RHS, and IBM tasks when compared to the verification case (fig. \ref{fig:barplot_perfo}a).

Additionally, fig. \ref{fig:samara_perfo}b displays the scaling of wall-time per time step across several domain resolutions and the two GPU devices (TITAN V and A100).
For the same domain size, the A100 halves the computational time compared to the TITAN V. 
For sufficiently large problems, both the TITAN V and the A100 show a linear scaling. 
However, the threshold for the linear scaling is about 7 million points for the TITAN V, and 20 million points for the A100. 
This effect is primarily due to GPU overheads such as under-utilization of GPU threads (i.e. Lagrangian GPU kernels), latency, memory allocation and kernels launch \cite{atomic}.
Only when the Eulerian and Lagrangian grids 
are sufficiently large
the GPU parallel processing capabilities can be fully leveraged, resulting in more efficient scaling and better performance.

Although not reported here, this study was also successfully performed in single precision, allowing the total number of points in the domain to double.
Indeed, the largest problem we were able to fit in the GPU using single precision had a size of roughly 650 million points (A100).

\subsection{Flow inside a simplified model of the left ventricle}
\begin{figure}[t]
    \centering
    \hspace*{-0.2cm}
    \includegraphics[width= 0.95\textwidth]{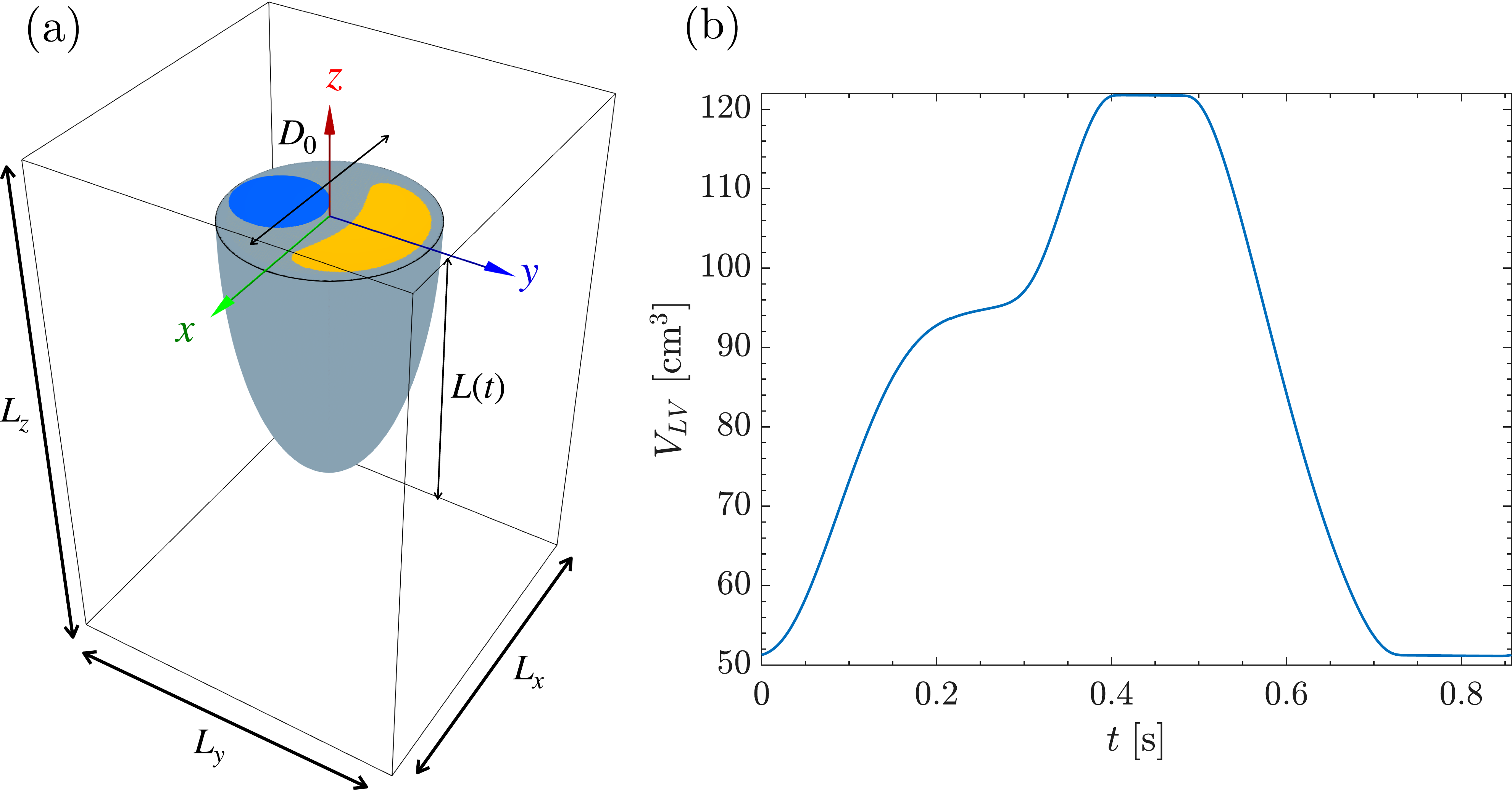}
    \caption{Modeled left ventricle geometry (a) and left ventricular volume in a time cycle (b).}
    \label{fig:LVV1}
\end{figure}

We employ TUCANGPU-3P to analyze the hemodynamics of a left ventricle.
For this simulation, the geometry is modelled as half of an ellipsoid with circular section of equation:
\begin{equation}
    \left(\frac{2x}{D_0}\right)^2+\left(\frac{2y}{D_0}\right)^2+\left(\frac{z}{L(t)}\right)^2 = 1,
\end{equation}
Here, $D_0$ represents the constant diameter of the plane of symmetry perpendicular to the $z$-direction with value $D_{0} = 5.46$ cm, and $L$ is a function of the simulation time $t$ designed to replicate the systole and diastole phases of the left ventricle volume.
Figure \ref{fig:LVV1}a provides a visual representation of the geometry of the left ventricle, highlighting these two parameters.
Throughout one cycle with a period of $t_{\text{cycle}} = 0.857$ s, the motion of the left ventricle's wall follows the temporal evolution of the parameter $L$. 
This parameter oscillates between $L_{\text{max}} = 7.803$ cm and $L_{\text{min}} = 3.276$ cm, to model the maximum expansion and contraction of the left ventricle.

We ran the simulation for 12 cycles, starting from a zero-flow initial condition. The kinematic viscosity was set to $\nu=0.04 \mathrm{cm^2/s}$, with a uniform grid spacing of $\Delta x = 0.03125$ cm in all directions, resulting in a grid size of $N_x \times N_y \times N_z = 256 \times 256 \times 512$ ($\sim 34$ millions of points). 
A constant time step was used to ensure $\text{CFL}< 0.3$ throughout the simulation.

The Mitral and Aortic valves serve as the interior inlet and outlet, respectively. They are simplified as planes that instantly open (when $Q>0$) and close (when $Q=0$) during the cardiac cycle. %both expansion and contraction.

The Aortic valve is represented by a circle with a radius $R_{A} = 2.516$ cm and center $y_A=-1.44$ cm, while the Mitral valve is described by a quartic curve equation:
\begin{equation} 
\left[ \left(\frac{2x}{D_0} \right)^2 + \left( \frac{2y}{D_0} \right)^2 - 1 \right]
\cdot 
\left[ \left( \frac{x}{R_{A}} \right)^2 + \left(\frac{y-y_{A}}{R_{A}} \right)^2 - 1\right] 
= \varphi,
\end{equation}
where the real number $\varphi$ is chosen in such a way that the desired Mitral valve area, $A_{MV} = 7.457$ cm$^2$, is ensured.

Figure \ref{fig:LVV1}b shows the temporal history of left ventricular volume in this particular case, highlighting the four distinct phases of the cardiac cycle.
The cycle starts with ventricular filling, occurring during the time interval $t \in [0, 0.4]$ s, coinciding with the opening of the Mitral valve and the subsequent expansion of the LV.
Following ventricular filling, the closure of the Mitral valve marks the end of diastole and initiates the systole phase. This phase spans from $t \in [0.4, 0.49]$ s.
The interval between the closure of the Mitral valve and the opening of the Aortic valve is referred to as the iso-volumetric contraction period, encompassing $t \in [0.4, 0.49]$ s.
Upon the opening of the Aortic valve, the left ventricle begins the rapid ventricular ejection phase, occurring within the time interval $t \in [0.49, 0.72]$ s. 
This phase continues until the left ventricle reaches its minimum volume.
Following the rapid ventricular ejection phase, the aortic valve closes. Then, with both valves closed, the iso-volumetric relaxation phase ensues, taking place between $t \in [0.72, 0.857]$ s. 

\begin{figure}[h]
    \centering
    \includegraphics[width= \textwidth]{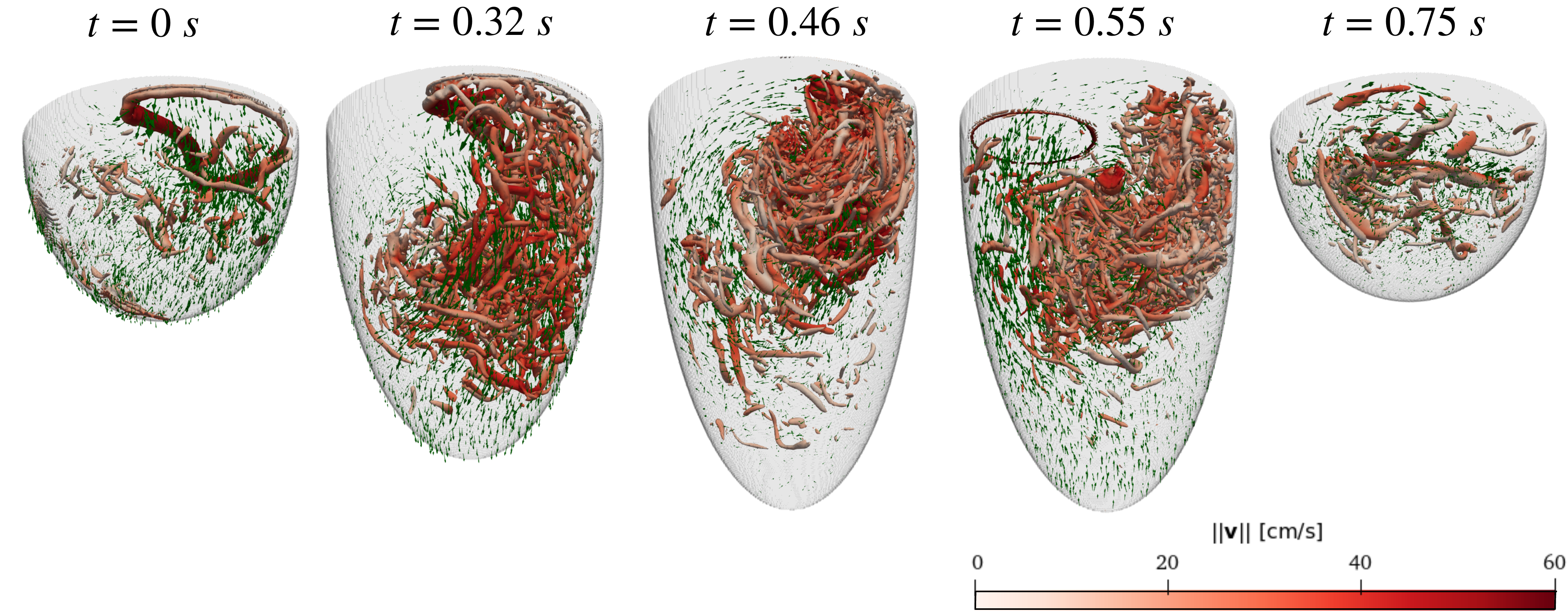}
    \caption{Visualization of the instantaneous velocity vector dield (depicted as green arrows) alongside iso-surfaces representing the second invariant of the velocity fradient tensor (with $Q = 10000 ; \mathrm{s^{-2}}$) color-coded to reflect the Local Velocity Magnitude $||\mathbf{v}||$.}
    \label{fig:paraview_LV}
\end{figure}

Figure \ref{fig:paraview_LV} provides a visual representation of 3D flow dynamics at five different time instants within the cardiac cycle. 
In these snapshots, we depict the velocity vector field together with iso-surfaces that illustrate the second invariant of the velocity gradient Tensor (Q-criterion).
At the beginning of the cardiac cycle ($t = 0$ s), the opening of the Mitral valve initiates the ventricular expansion, leading to an increase in inlet velocity. 
This expansion promotes the generation of vortices along the edges of the Mitral valve.
These vortices are subsequently transported towards the apex of the ventricle, favoring the mixing of blood, which becomes particularly evident at $t = 0.32$ s.
Following the closure of the Mitral valve and the cessation of ventricular expansion, a clockwise recirculation pattern emerges, resulting in the accumulation of vortices in the vicinity of the Mitral valve $t = 0.46$ s.
The opening of the Aortic valve marks the beginning of the ejection phase. 
At $t = 0.55$ s, the clockwise recirculation pattern is enhanced, leading to an increase in the velocity at the region surrounding the Aortic valve.
This clockwise flow pattern expels blood at a high velocity as the ventricle contracts, ultimately reaching its minimum volume, as observed at $t = 0.75$ s.

\begin{figure}
    \centering
    \includegraphics[width= 1\textwidth]{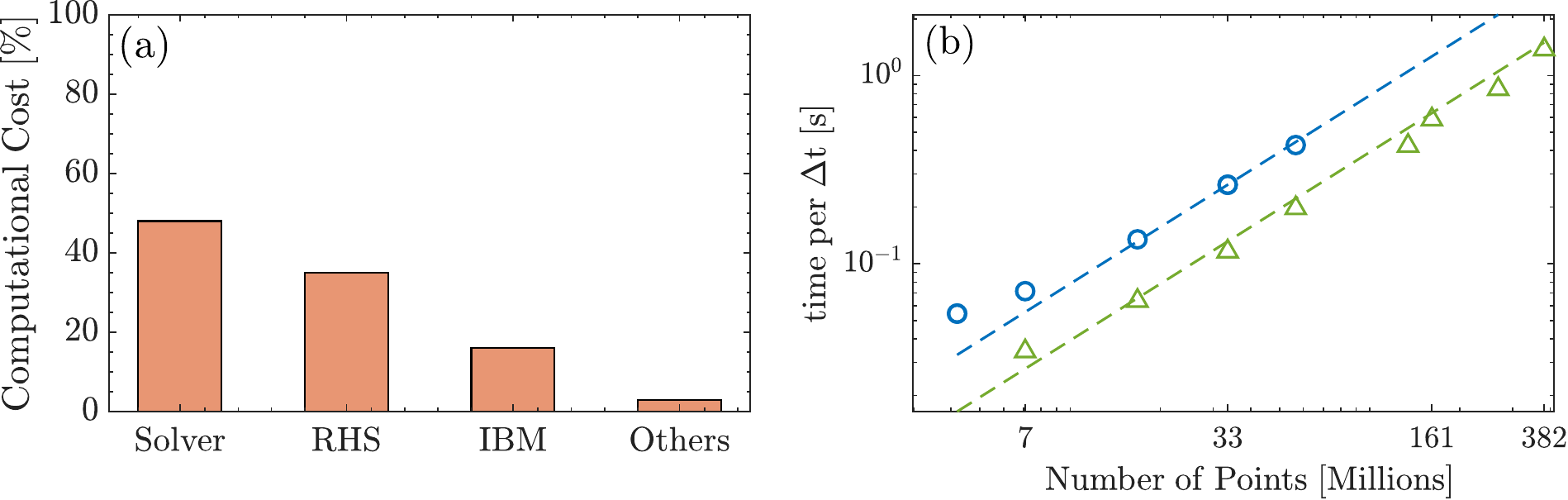}
    \caption{Performance indicators in the left ventricle problem. (a) Bar plots for the relative contribution of different tasks performed on the GPU to the total computational cost per time step. (b) The wall-time per $\Delta t$ is displayed for different domain resolutions. Dashed lines represent the ideal strong scaling with domain size for two different GPU hardware: TITAN V (\protect\dashedclearblueline) and A100 (\protect\dasheddarkergreenline).}
    \label{fig:LV_perfo}
\end{figure}

Finally, we quantify the distribution of computational costs and evaluate the performance of the GPU code for the left ventricular flow simulation using both GPU cards. %(TITAN V and A100). % hardware TITAN V and A100.
Figure \ref{fig:LV_perfo}a depicts the relative contribution of the GPU-performed tasks to the total computational cost per time step. 
The discretization of the immersed surfaces in cardiovascular flows demands considerably larger amounts of Lagrangian points compared to the external aerodynamics simulations. 
Indeed, fig. \ref{fig:LV_perfo}a shows a notable rise in the relative time spent in IBM interpolations for the left ventricle simulation compared to the verification cases (fig. \ref{fig:barplot_perfo}b), as a consequence of the increased ratio between Lagrangian and Eulerian points.
The relative computational costs dedicated to the remaining tasks (Solver and RHS), on the other hand, remain similar.
Additionally, fig. \ref{fig:LV_perfo}b demonstrates proper strong scaling across various domain resolutions, after a threshold of roughly $7$ million points for both GPUs.
Compared to the bioinspired aerodynamics case (fig. \ref{fig:samara_perfo}), the linear scaling starts at the same number of points for the A100 and the TITAN V, suggesting that the Lagrangian mesh in the LV simulations is large enough to avoid the under-utilization of the A100 in the Lagrangian kernels. 
As before, this study was successfully tackled in single precision, allowing for the simulation of a domain with up to approximately 760 million points for the A100.

%%%%%%%%%%%%%%%%%%%%%%%%%%%%%%%%%%%%%%%%%%%%%%%%%%%%%%%%%%%%%
%%%%%%%%%%%% Conclusions
%%%%%%%%%%%%%%%%%%%%%%%%%%%%%%%%%%%%%%%%%%%%%%%%%%%%%%%%%%%%%
\section{Conclusions}\label{sec:conclusions}
%%%%%%%%%%%%%%%%%%%%%%%%%
%%%%context%%%%%%%%%%%%%%

The latest advancements in GPU technology, including massive parallelization, increased bandwidth, and expanded memory capacity, have led to the development of accessible programming languages such as CUDA and highly optimized libraries like CuFFT for Fourier transforms \cite{cuda}. 
These developments have simplified GPU integration into computational workflows, motivating researchers to adapt CPU-based solvers for efficient GPU acceleration. 
This is particularly relevant in the CFD community, where DNS at moderately low Reynolds numbers have broad applications, from bio-inspired aerodynamics to cardiovascular flows.
These complex problems often involve intricate moving geometries that can be modeled efficiently using the Immersed Boundary Method, decoupling fluid and immersed body meshes. 
Historically, such simulations required substantial computational resources, leading to the development of parallel CPU architectures for high-performance computing clusters. 
Using GPUs suited for massive parallel computations, ported codes to GPU architectures now offer unprecedented speed-ups.

In this context, we introduce TUCANGPU, a Python-based code offering a numerical framework tailored for DNS of incompressible flows. 
Evolved from its CPU version TUCAN, a MPI-based parallel code written in Fortran, TUCANGPU leverages the exponential growth in GPU memory capacity to perform simulations with up to 380 million grid-points (double precision in the A100).

We redesigned the parallelization algorithm to optimize TUCANGPU for efficient operation on a single GPU. 
This approach eliminates the need for CPU-CPU communications and reduces CPU-GPU data transfer, simplifying the computational process and significantly avoiding bottlenecks for more effective simulations. % of complex, moving geometries submerged in a fluid. %in various scientific and engineering fields.
In addition, the data transfer between the different hierarchical types of memory in the GPU (i.e., global, shared, and local) is optimized by tailoring the GPU grid structure to each operation, evenly distributing the workload among the different GPU threads.
TUCANGPU also leverages GPU-accelerated libraries provided by CuPy to handle the most computationally intensive tasks, which are the solution of the linear systems derived from the Poisson and Helmholtz equations.

Taking all these into consideration, the numerical algorithm is implemented with the following design principles: firstly, the CPU is responsible solely for I/O operations. Three types of GPU grids are designed to evenly distribute Eulerian, Lagrangian, and LU factorization operations. 
Helmholtz and Poisson equations are directly solved using Fourier transforms (FFTs) from CuPy in the periodic directions and LU matrix inversion in the non-periodic direction. 
This approach allows for the direct solution of the linear systems in the numerical algorithm, contrasting with the CPU version, which solves the linear systems iteratively using HYPRE.

After presenting several verification benchmarks, we compare the results obtained by TUCANGPU and TUCAN across different flow configurations. 
We then assess the solver performance by examining key tasks in the algorithm of both codes: solution of the linear systems, computing the right-hand sides, and performing interpolations required by the immersed boundary method (IBM).

The benchmark shows a significant speed-up in the solution of the linear systems respect to its CPU version, ranging from a factor of 30 to 80 for low and high resolutions, respectively. 
Additionally, optimizing the IBM interpolations by avoiding communication between processors and parallelizing by Lagrangian points results in a speed-up of 7 to 12 depending on the domain resolution.

Finally, to evaluate the variability in performance of TUCANGPU, we report two applications with significantly different proportions of Lagrangian and Eulerian points, representative of external and internal flows.
For the external flow application, we study a rotating winged-seed. 
In this scenario, the region of interest is much larger than the immersed body, resulting in a small ratio between Lagrangian and Eulerian points. On the other hand, for the internal flow application, we consider hemodynamics in a simplified left ventricle. 
Here, the size of the immersed body is comparable to the fluid domain, leading to a proportionally larger discretized surface and a higher ratio of Lagrangian to Eulerian points.
In this study, we measure the strong scaling of TUCANGPU using two different GPU cards with different specifications: the TITAN V and the A100. 
The software demonstrates effective scalability in both external and internal flow setups with domain resolutions starting from approximately 7 million points.
The maximum problem size that the TITAN V could fit in double precision was approximately 60 million points, achieving a wall-time per time step of around 0.5 seconds. 
In comparison, the A100 doubled the performance of the TITAN V for the same domain size and could handle a maximum problem size of 380 million points in double precision thanks to its larger memory, achieving a wall-time per time-step of 1.37 seconds.

While the implementation presented in this work showcases some limitations in terms of memory restrictions and periodicity in the boundary conditions, in practice this study demonstrates the effectiveness of modern GPU technology to successfully tackle moderately large  numerical simulations for a wide variety of flow configurations of interest, even in commodity hardware.

\section*{Acknowledgements}

This work was partially supported by grants PID2019-107279RB-I00 and TED2021-131282B-I00, both funded by the \sloppy{MCIN/AEI/10.13039/501100011033} and the European Union NextGenerationEU/PRTR, and by grant 1R01HL160024 from the National Institutes of Health. 

\section*{Declaration of interests}
The authors report no conflict of interest.
%%%%%%%%%%%%%%%%%%%%%%%%%%%%%%%%%%%%%%%%%%%%%%%%%%%%%%%%%%%%%
%%%%%%%%%%%%%%%%%%%%%%%%%%%%%%%%%%%%%%%%%%%%%%%%%%%%%%%%%%%%%
%%%%%%%%%%%%%%%%%%%%%%%%%%%%%%%%%%%%%%%%%%%%%%%%%%%%%%%%%%%%%
%%%%%%%%%%%%%%%%% Appendices
%%%%%%%%%%%%%%%%%%%%%%%%%%%%%%%%%%%%%%%%%%%%%%%%%%%%%%%%%%%%%
%%%%%%%%%%%%%%%%%%%%%%%%%%%%%%%%%%%%%%%%%%%%%%%%%%%%%%%%%%%%%
%%%%%%%%%%%%%%%%%%%%%%%%%%%%%%%%%%%%%%%%%%%%%%%%%%%%%%%%%%%%%
\appendix
%%%%%%%%%%%%%%%%%%%%%%%%%%%%%%%%%%%%%%%%%%%%%%%%%%%%%%%%%%%%%
%%%%%%%%%%%% Details on the tri-diagonal systems solution
%%%%%%%%%%%%%%%%%%%%%%%%%%%%%%%%%%%%%%%%%%%%%%%%%%%%%%%%%%%%%
\section{Modified wavenumbers definition}\label{sec:wavenumber}

Let us consider the second spatial derivative of a generic variable $\Theta$, approximated at point $x_i$ using a central finite difference approximation of second order:
\begin{equation} \label{eq:wavenumber1}
    \frac{\partial^2 \Theta}{\partial x^2} \biggr\rvert_i = \frac{\Theta_{i+1} - 2\Theta_i + \Theta_{i-1}}{h^2} + \mathcal{O}(h^2).
\end{equation}
If $\Theta$ is periodic along the $x$-direction, we can express $\Theta_i$ as $\Theta_i=\widehat{\Theta_n} \exp(I \kappa_n i h)$ using Fourier, where $\kappa_n$ is the $n$-th wavenumber out of $N$ and $I$ is the imaginary unit.
We can then compute the second derivative easily:

\begin{equation} \label{eq:wavenumber2}
 \frac{\partial^2 \Theta}{\partial x^2} \biggr\rvert_i = - \kappa_n^2  \widehat{\Theta_n} \exp(I \kappa_n i h) + \mathcal{O}(h^N).
\end{equation}
Considering the \textit{modified wavenumber} $\kappa_n'$ a second-order derivative can be expressed as:
\begin{equation} \label{eq:wavenumber2.5}
 \frac{\partial^2 \Theta}{\partial x^2} \biggr\rvert_i = - \kappa_n'^2  \widehat{\Theta_n} \exp(I \kappa_n i h) + \mathcal{O}(h^2).
\end{equation}
Equations \eqref{eq:wavenumber1} and \eqref{eq:wavenumber2.5} can be made compatible by expressing the terms $\Theta_{i+1}, \Theta_i$ and $\Theta_{i-1}$ as a Fourier series, i.e. $\Theta_{i\pm 1}=\widehat{\Theta_n} \exp[I \kappa_n (i\pm 1) h]$.
Then eq. \eqref{eq:wavenumber1}, after dropping the higher order terms, becomes:
\begin{equation} \label{eq:wavenumber3}
    \frac{\partial^2 \Theta}{\partial x^2} \biggr\rvert_i \simeq \widehat{\Theta_n} \exp(I \kappa_n i h)  \frac{\exp(-I\kappa_nh)-2+\exp(I\kappa_nh)}{h^2}.
\end{equation}

We can thus infer an expression for the \textit{modified wavenumber} $\kappa_n'$, from eqs. \eqref{eq:wavenumber2.5} and \eqref{eq:wavenumber3}, such that:
\begin{equation}\label{eq:wavenumber4}
-\kappa_n'^2 = \frac{\exp(-I\kappa_nh)-2+\exp(I\kappa_nh)}{h^2},
\end{equation}
that after simplifying yields:
\begin{equation}\label{eq:wavenumber5}
\kappa_n'^2 = \frac{2}{h^2} [1-\cos(\kappa_n h)].
\end{equation}

In this work, we extend this concept to the three dimensions, expressing the Laplacian operator of eqs. \eqref{rksubstep:e} and \eqref{rksubstep:f} with their corresponding modified wavenumbers in the periodic directions.
For clarity, the expression for the corresponding vector of modified wavenumbers employed in TUCANGPU-3P is:
\begin{equation}
    \kappa'^2_{ijk}= \frac{2}{h^2}[3 -\cos(\kappa_{x_i} h) - \cos(\kappa_{y_j} h) - \cos(\kappa_{z_k} h)],
\end{equation}
where $\kappa_x, \kappa_y$ and $\kappa_z$ are the wavenumbers along the $x,y$ and $z$ directions respectively. 
For the two-periodic case (TUCANGPU-2P), the vector of modified wavenumbers can be obtained  accordingly:
\begin{equation}
    \kappa'^2_{jk}= \frac{2}{h^2}[2 - \cos(\kappa_{y_j} h) - \cos(\kappa_{z_k} h)].
\end{equation}

\section{Details on the boundary conditions on the tri-diagonal systems} 
\label{sec:appendix_linearsystems}
We provide some details on the tri-diagonal systems that need to be solved in TUCANGPU-2P.
For the intermediate velocity components, eq. \eqref{eq:tridiagsys} defines the coefficients of each linear system as:
\begin{subequations}\label{eq:coeff}
    \begin{equation}
      A_1 = - \left[\frac{2}{h^2}+ \kappa_y'^2+ \kappa_z'^2+ \frac{\text{Re}}{\Delta t \beta_n} \right], \quad 
      A_2 = \frac{1}{h^2}.%, \quad
     \end{equation}
\end{subequations}
We provide the tri-diagonal system for the $jk$-th pencil, for the modes of the intermediate velocity, $\widehat{\mathbf u^*}$, considering Dirichlet conditions at both boundaries:
\begin{equation}
\begin{pmatrix}
 A_{11} &  A_2    & 0      & 0      & 0      & 0      & 0      \\ 
 A_2    &  A_1    & A_2    & 0      & 0      & 0      & 0      \\ 
 0      &  A_2    & A_1    & A_2    & 0      & 0      & 0      \\ 
 0      &  0      & A_2    & A_1    & A_2    & 0      & 0      \\ 
 \vdots &  \vdots & \vdots & \ddots & \ddots & \ddots & \vdots \\ 
 0      &  0      & 0      & 0      & A_2    & A_1    & A_2    \\ 
 0      &  0      & 0      & 0      & 0      & A_2    & A_{11}  \\ 
\end{pmatrix}
_{j,k}
\begin{pmatrix}
\widehat{\mathbf u^*}_2  \\ 
\widehat{\mathbf u^*}_3  \\ 
\widehat{\mathbf u^*}_4  \\ 
\widehat{\mathbf u^*}_5  \\ 
\vdots   \\ 
\widehat{\mathbf u^*}_{N_x-2} \\
\widehat{\mathbf u^*}_{N_x-1} \\ 
\end{pmatrix}
_{j,k}
= 
\begin{pmatrix}
\widehat{\mathbf{RHS}}_2 - \alpha A_2\widehat{\textbf{BC}}_{u,1}  \\ 
\widehat{\mathbf{RHS}}_3   \\ 
\widehat{\mathbf{RHS}}_4    \\ 
\widehat{\mathbf{RHS}}_5    \\ 
\vdots   \\ 
\widehat{\mathbf{RHS}}_{N_x-2} \\
\widehat{\mathbf{RHS}}_{N_x-1}  - \alpha A_2 \widehat{\textbf{BC}}_{u,N_x}    \\  
\end{pmatrix}
_{j,k},
\end{equation}
where $\alpha=1$ for the velocity component in the non-periodic direction, and $\alpha=2$ for the staggered directions velocity components, for which the boundary is not aligned with the grid points, and $A_{11}=A_1-A_2(\alpha-1)$.
We proceed similarly for the pseudo-pressure modes, $\widehat \phi$, using Neumann boundary conditions.
We define a new coefficient $A_3$ that will go in the diagonal:
\begin{equation}\label{eq:coeff2}
     A_3 = -\left[\frac{2}{h^2}+ \kappa_y'^2+ \kappa_z'^2 \right].
\end{equation}
The system of equations for each pseudo-pressure mode $\widehat{\phi}_{jk}$ is then:
\begin{equation}
\begin{pmatrix}
 A_3+A_2&  A_2    & 0      & 0      & 0      & 0      & 0      \\ 
 A_2    &  A_3    & A_2    & 0      & 0      & 0      & 0      \\ 
 0      &  A_2    & A_3    & A_2    & 0      & 0      & 0      \\ 
 0      &  0      & A_2    & A_3    & A_2    & 0      & 0      \\ 
 \vdots &  \vdots & \vdots & \ddots & \ddots & \ddots & \vdots \\ 
 0      &  0      & 0      & 0      & A_2    & A_3    & A_2    \\ 
 0      &  0      & 0      & 0      & 0      & A_2    & A_3+A_2  \\ 
\end{pmatrix}
_{j,k}
\begin{pmatrix}
\widehat{\phi^*}_2  \\ 
\widehat{\phi^*}_3  \\ 
\widehat{\phi^*}_4  \\ 
\widehat{\phi^*}_5  \\ 
\vdots   \\ 
\widehat{\phi^*}_{N_x-2} \\
\widehat{\phi^*}_{N_x-1} \\ 
\end{pmatrix}
_{j,k}
= 
\begin{pmatrix}
\widehat{\mathrm{RHS}}_{\phi,2} + A_2 h \widehat{\mathrm{BC}}_{\phi,1}  \\ 
\widehat{\mathrm{RHS}}_{\phi,3}   \\ 
\widehat{\mathrm{RHS}}_{\phi,4}    \\ 
\widehat{\mathrm{RHS}}_{\phi,5}    \\ 
\vdots   \\ 
\widehat{\mathrm{RHS}}_{\phi,N_x-2} \\
\widehat{\mathrm{RHS}}_{\phi,N_x-1}  - A_2 h \widehat{\mathrm{BC}}_{\phi,N_x}     \\  
\end{pmatrix}
_{j,k}.
\end{equation}

Note that employing homogeneous Neumann boundary conditions, i.e. $\partial \phi/\partial x = 0$ at the non-periodic boundaries, is equivalent to setting $\widehat{\mathrm{BC}}_{\phi,1}=\widehat{\mathrm{BC}}_{\phi,N_x}=0$.
%
%Also, we account for the singularity that arises when imposing Neumann boundary conditions as explained in section \ref{sec:inflow-outflow}.
%
Also, since the grid is uniform, we avoid ill-conditioning by damping all the high-frequency modes, keeping the second-order of accuracy of the scheme even in the periodic directions.
This is accomplished using modified wavenumbers as presented in appendix \ref{sec:wavenumber}.

\clearpage

%\bibliography{reference}
%\bibliographystyle{plainnat}
%\bibliographystyle{plain}
%\bibliographystyle{ieeetr}

\end{document}